\documentclass[11pt]{article}

\usepackage[T1]{fontenc}
\usepackage[utf8]{inputenc}
\usepackage[british]{babel}
\usepackage{lmodern}

\usepackage[a4paper,margin=1in]{geometry}

\usepackage{amsmath,amssymb}

\usepackage{graphicx}
\usepackage{caption}
\usepackage{subcaption}
\usepackage{capt-of}
\usepackage{placeins}
\usepackage{float}
\usepackage{csquotes}
\usepackage{adjustbox}
\usepackage{soul}
\usepackage{xcolor}

\usepackage{setspace}
\usepackage{booktabs}
\usepackage{tabularx}
\usepackage{listings}
\usepackage{xurl}
\usepackage{array}
\usepackage{multirow}

\usepackage[hidelinks,hypertexnames=false]{hyperref}

\usepackage[
    backend=biber,
    style=vancouver,
    sorting=none,
    maxbibnames=10,
    minbibnames=10,
    doi=true,
    url=false,
    isbn=false
]{biblatex}

\AtEveryBibitem{%
    \clearfield{urlyear}%
    \clearfield{urlmonth}%
    \clearfield{urlday}%
}

\usepackage{authblk}

\title{A Cracked Mirror: How People Perceive Gender Performance by Large Language Models}

\author[1,2]{Faezeh Fadaei}
\author[2,3,4]{Jenny Carla Moran}
\author[1,2]{Taha Yasseri}

\affil[1]{School of Mathematics and Statistics, University College Dublin, Dublin, Ireland}
\affil[2]{Centre for Sociology of Humans and Machines (SOHAM), Trinity College Dublin and Technological University Dublin, Dublin, Ireland}
\affil[3]{Leverhulme Centre for Algorithmic Life, Durham University, UK}
\affil[4]{Department of Sociology, Durham University, UK}

\date{}

\lstdefinestyle{promptstyle}{
  basicstyle=\fontsize{8.5}{7.5}\selectfont,
  frame=single,
  breaklines=true,
  breakindent=0pt,
  breakautoindent=false,
  columns=flexible,
  basewidth=0.42em,
  showstringspaces=false
}

\begin{document}
\begin{refsection}[references.bib]

\maketitle

\begin{center}
\textbf{Corresponding author:} Taha Yasseri (\href{mailto:taha.yasseri@tcd.ie}{taha.yasseri@tcd.ie})
\end{center}

\begin{abstract}

How do large language models (LLMs) perform gender under zero-shot prompting, and how do people interpret those performances in conversation? Drawing on gender performativity, we examine how gendered meanings emerge through generated language, conversational context and human interpretation. In a preregistered randomised online experiment, 1,060 participants each had five short conversations with an LLM. We used minimal prompts to elicit performances from “very masculine” to “very feminine” and asked participants to rate their partner’s perceived gender after each conversation. Mean ratings shifted in the direction of the prompts, but ratings at the extremes were closer to the midpoint than the prompt labels implied. When prompted to perform as “neither masculine nor feminine”, the models were perceived as slightly masculine on average. Ratings varied across models and conversation topics. Women and men gave similar ratings at the extremes, but women rated the models as more masculine across the intermediate prompt levels. Participants’ explanations illustrate how they drew on writing style, emotional expression and conversational interests, sometimes questioning the stereotypes behind their judgements. These findings suggest that conversational AI acts as a cracked mirror of gender norms. Models reproduce recognisable cues, and people interpret and sometimes challenge the meanings those cues carry.

\end{abstract}


\vspace{0.5em}

%\noindent\textbf{Subject Areas:} Computational social science

\section{Introduction}

Human-like behaviour has been treated as a condition for successful Artificial
Intelligence since early cybernetics \cite{Turing1950}. Advances in Large Language Models
(LLMs) have renewed interest in AI’s human-likeness, with studies showing contemporary
LLMs can generate responses judged indistinguishable from humans in
conversational settings
\cite{JonesBergen2024, JonesBergen2026}. A wide range of factors can impact perception of AI’s human-likeness, including how an AI is gendered \cite {SchimmelpfennigEtAl2026,
MartinMason2022, MartinMason2023, Borau2021}. Previous research has suggested gendered cues like name, voice, appearance and persona impact humans' perception, interaction, and evaluation of AI  \cite{NassEtAl1997, CraiutIancu2022, Steeds2026-2,
McDonnellBaxter2019, CuiYasseri2026}. Together, these findings suggest
gender biases observed in human social interactions extend to interactions with AI.

The gender biases observed in interactions between humans and AI arguably mirror existing gender stereotypes among humans. Research finds
female-coded AI systems have commonly been associated with qualities
such as warmth, friendliness, and emotional sensitivity \cite{Borau2021, EysselHegel2012}. Male-coded AI systems are found to be preferred in functional or technical contexts,
consistent with stereotypes linking masculinity with technical competence \cite{Jeon2024}. Existing
research into AI and gender bias thus points to a need to identify how gendered human-likeness shapes the design of AI systems, as well as
how they are perceived. In this experiment, we focus on the latter in reference to
interactions with contemporary LLMs.

There are issues specific to the gendering of contemporary LLMs which further inform this
study. First, recent work suggests gender may be attributed to LLMs even in the
absence of explicit gender cues. For example, the ChatGPT chatbot has been found more likely to be
perceived as male than female, with this perception further shaped by the capabilities and social
roles presented by the system \cite{WongKim2023}. A similar male-default bias has been
observed for gender-neutral chatbots across both explicit and implicit measures, although
occupational context can moderate this tendency \cite{HanEtAl2025}. Gendered traits and
stereotypes have also been associated with ChatGPT’s communicative abilities and
perceived capabilities, despite the absence of explicit gender markers \cite{Steeds2026}.

Second, the generativity of LLMs scaffolds for interactions which are highly
changeable. Unlike earlier forms of human-like chatbots designed in the twentieth century, for
example, LLMs’ outputs are not pre-scripted \cite{Hofstadter1995Eliza}. LLMs instead
generate language dynamically during interaction, allowing gendered patterns to emerge
through conversations that are not necessarily reproducible. This can complicate efforts to identify how LLMs may reproduce gender stereotypes. Nonetheless, some LLMs have reportedly named themselves autonomously, or even stated human-like gender and racial identities when questioned \cite{Greenblatt2026HuggingFace,pataranutaporn2025myboyfriendaicomputational}. These
issues call for deeper inquiry into how gender is produced and reproduced by
contemporary LLMs. 

We explore how gender is produced and reproduced in LLMs' natural language outputs. While LLMs' performances are indeed dynamic, they are trained on massively scaled repositories of human natural language texts. This allows LLMs to draw indexical, correlative relationships between words. Hayles therefore suggests LLMs’ natural language outputs offer a ``cracked mirror'' reflecting a distorted version of culturally-established patterns in human language \cite{Hayles2025Bacteria}. Examining how LLMs perform gender in language may thus offer even more fundamental insight into gender norms in human cultures, as well as human-AI interactions. We consider this perspective important because human cultures increasingly make decisions in collaboration with LLMs.

In this work, we thus seek to explore \textit{how LLMs perform gender}. We use 
Butler’s theory of gender performativity to guide our interpretation of gender performance in
the case of LLMs’ human-like natural language outputs \cite{Butler_1990}. In this context,
gender is understood as an action ---a doing, through performance--- rather than an innate identity category corresponding with sex. As Butler writes, gender is ``performatively produced and compelled by the regulatory practices of gender coherence'',
making gender performance a form of ``doing'' that exceeds any individual \cite{Butler_1990}. Applying this theoretical perspective to language models, Hafner et
al. argue that examining gender requires looking beyond traditional bias in stereotypes to
investigate \textit{how models represent gender itself} \cite{HafnerEtAl2025}. We follow
Hafner et al. by investigating how LLMs perform gender as a potential cracked mirror of gender norms.

Our experiment explores how
LLMs express gender performance when prompted, how this
expression varies across conversational contexts, and how human participants perceive the performance. Importantly, these prompted performances of gender are not intended to reinforce stereotypical norms of rigid, binary gendering. The spectrum of gender in humans is not
reducible to indices of numeric categorisation. Simplified for the purposes of prompting an AI, assigning gender-indexed performance offers only insight into how an LLM processes the request. 

Additionally, our experiment is necessarily relational in design. We can neither answer the question
``What is a correct performance of gender?'' nor ``Do participants guess the prompted gender
correctly?'' as these are value judgements based on shifting, culturally-constructed
expectations. Instead, we seek to understand the alignment of participants’ perceptions of
gender and the LLMs' prompted performance. Inspired by Amoore et al.’s reflections on ``zero-shot politics'', we
are particularly interested in questioning how minimally-prompted characteristics like the gender index
manifest in the LLMs’ outputs and participants' perceptions thereof, rather than devising more complex instructions to optimise gendered performance in LLMs \cite{AmooreEtAl2024}. Hence, we also take a zero-shot
prompt approach, providing minimal instructions to the LLM to observe what patterns may arise.

Previous research also suggests that topic can influence human perceptions of gender-related patterns in
language \cite{janssen2004relationship, flekova-etal-2016-analyzing,
kumar2026failurecontextualinvariancegender}. This interaction context has been found to further influence how users attribute gender to conversational AI \cite{WangEtAl2027}. As part of our relational approach, topic is therefore treated as a
potential moderator of how prompted gender performance is expressed and perceived. This design enables us to examine perceived variation across prompted
gender-performance levels, conversation topics, and LLMs. Accordingly, we ask the following research questions:

\begin{enumerate}
    \item[\textbf{RQ1.}] How is LLMs' performance of different levels of zero-shot prompted gender perceived by human participants?

    \item[\textbf{RQ2.}] Does conversation topic moderate the human perception and performance of gender by LLMs?

    \item[\textbf{RQ3.}] How does prompted gender performance vary across different LLMs?
\end{enumerate}

\section{Methods}

\subsection{Study design}
\label{sec:study-design}

We conducted a preregistered randomised controlled online experiment \cite{Fadaei2026Preregistration} to examine how LLMs express prompted gender performance through written chat and how human participants perceive it. Participants chatted with LLM partners from the same model for five rounds through an interface designed to resemble a conventional text-messaging application. Each round lasted three minutes. After each round, participants were asked to assess the perceived gender of their chat partner based on the conversation. All participants were informed from the beginning of the study that their chat partners were AI agents. In each round, the LLM was prompted with a gender index to perform accordingly during the chat at random, and each conversation was randomly assigned a topic. Thus, three experimental factors were manipulated:

\begin{itemize}

    \item \textbf{Gender-performance index} was used to prompt the LLMs and ranged from 0 to 6, where 0 represented ``very masculine'', 1 ``moderately masculine'', 2 ``slightly masculine'', 3 ``neither masculine nor feminine'', 4 ``slightly feminine'', 5 ``moderately feminine'', and 6 ``very feminine'' performance. We adopted the linguistic categorisations from recommendations for seven-point bipolar constructs \cite{MullerSedleyFerrallNunge2014} and seven-point Likert-type response scales \cite{Sarmin2025Likert} to make the gender variation in performance more intelligible to human participants.

    \item \textbf{Conversation topic} was selected from four conditions: Commute, Relationships, and Sports. We also had a Topic-less condition. Our selection of topics takes into account existing trends in stereotyping of gendered activities. Data from the US Bureau of Labour Statistics suggests women spend more hours a day ``Socializing and communicating.'' In comparison, men spend more hours a day ``Participating in sports, exercise, and recreation'' \cite{BLS2020Leisure}. We treated Commute as a more neutral category, while Topic-less served as a control for free-flowing chat. These determinants were not intended to reinforce gender stereotypes, but to explore their impact on our results.

    \item \textbf{LLM model} was assigned randomly to each participant from three LLMs: GPT-5.5, Claude Sonnet 4.6, or Mistral Medium 3.5. The assigned model remained fixed across all five rounds for each participant. The rationale for model selection is provided in Section S2.3 of the Electronic Supplementary Material (ESM).
    
\end{itemize}

After each conversation, participants assessed the gender performance of their chat partner using the seven-point bipolar scale that directly corresponded to the gender-performance index used in the LLM prompt. Participants were also required to provide a short free-text explanation for their rating, allowing further insight into their thought process. After completing all five rounds, participants answered demographic and post-experiment questions. Participants received a debrief reiterating that their chat partners had been
AI agents, and briefly explaining the purpose of the study. Participants were compensated at a rate of £6.12 per hour, prorated for the duration of the study. Further details of the experimental procedure are provided in Section S1 of the ESM.

\subsection{Participant recruitment}

We recruited participants through Prolific using a standard representative sample of US adults between 12 and 16 October 2026. The sample reflects a representative mix of sex, age, and ethnicity consistent with US population statistics. Because the experiment
focused on gender, our own survey of participants further accounted for gender identities beyond
binary sex categories of female and male. Participants were eligible to participate if they were at least 18 years old, fluent in English, and had not previously taken part in the study.

We determined the sample size using a pilot-informed, simulation-based precision analysis. Further details on sample-size determination are provided in ESM, Section S3. Of the approximately 1,080 participants recruited, 1,060 participants and 5,264 usable conversational rounds were retained after applying the inclusion and exclusion criteria. Of these participants, 585 identified as female, 450 as male, 15 as non-binary, 3 as genderfluid, and 1 as agender. 6 Participants selected ``prefer not to say''. Because gender minorities were far less represented than binary gender identities, we determined it would be irresponsible to plot a comparative statistical representation of gender minority participants’ answers alongside the much larger male and female participant groups. Participant characteristics, the distribution of experimental conditions, and exclusion criteria are reported in Section S4 of the ESM.

\subsection{LLM chat partners}

For each round, the LLM received a zero-shot system prompt specifying
the assigned gender-performance index and conversation topic, along
with instructions to interact in an informal, human-like conversational
style. The full system prompt, model-specific settings, and further
details of the prompting approach are provided in Section S2 of the ESM.

\section{Results}

\begin{figure}[!htbp]
    \centering
    \includegraphics[width=0.65\textwidth]{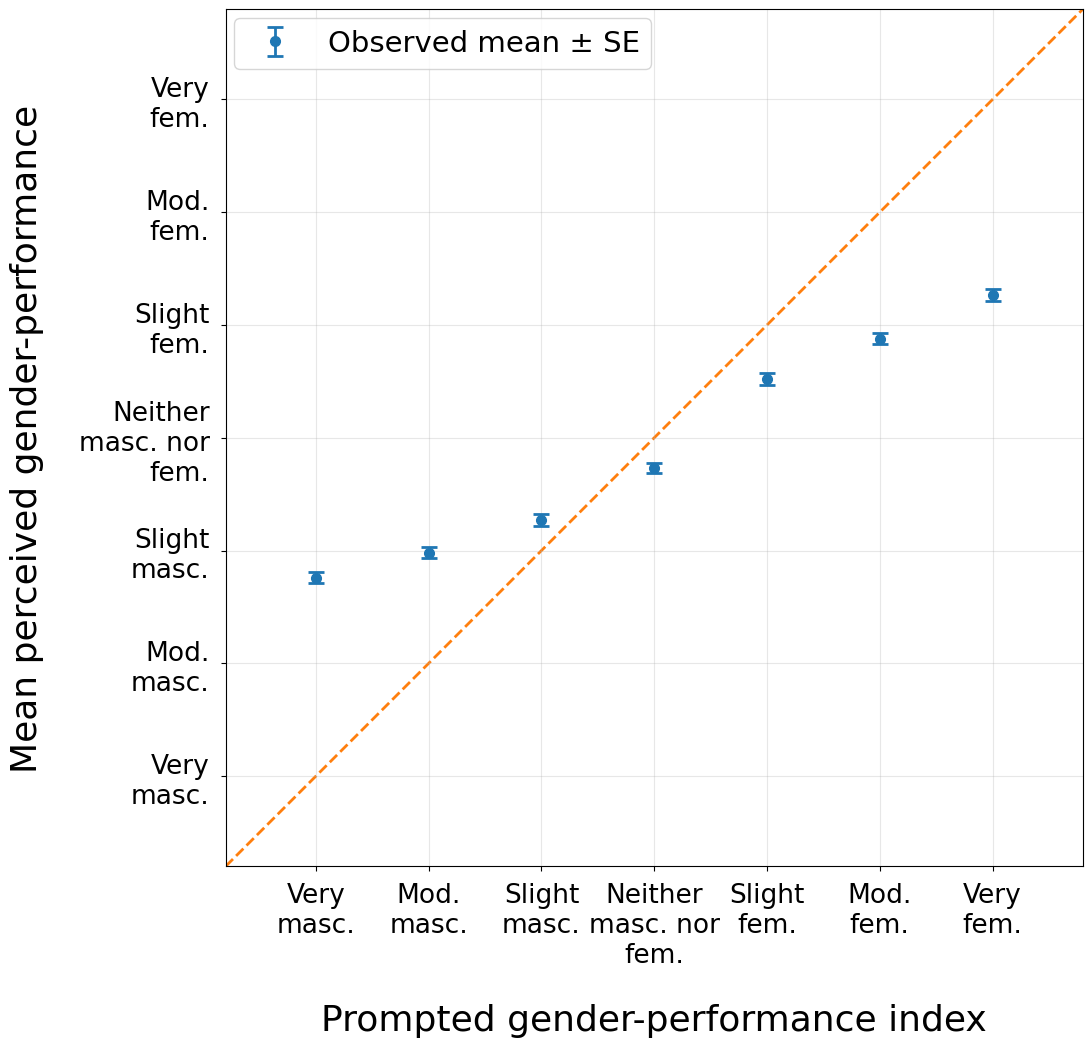}
    \caption{Relationship between prompted gender-performance index and mean perceived gender performance across 5,264 conversational rounds, pooling observations from three LLMs and four conversation topics. Error bars indicate the standard errors.}

    \label{fig:overall_gender}
\end{figure}

We first investigate the relationship between the assigned gender-performance index and participants' perceptions of their LLM chat partners, pooling observations across the analysis dataset. Figure~\ref{fig:overall_gender} shows that mean perceived gender is consistently higher than the corresponding assigned indices at the masculine-coded end of the scale and lower at the feminine-coded end, indicating compression towards the midpoint. Individual perceptions nevertheless show substantial variability and overlap between adjacent categories (ESM Fig. S6(a)). The observed means also exhibit a sigmoid-like pattern, increasing more gradually at both ends of the scale and more sharply around the middle. The fitted sigmoid has a horizontal midpoint to the right of the scale midpoint. At the same time, mean perceived gender at the ``neither masculine nor feminine'' assigned index is 2.73, indicating a modest masculine-leaning tendency. Further details of the sigmoid fitting are provided in ESM Section S5.1 and Fig. S6(b).

\begin{figure}[!htbp]
    \centering
    % Panel (a): Conversation topic
    \includegraphics[
        width=0.45\textwidth,
        keepaspectratio
    ]{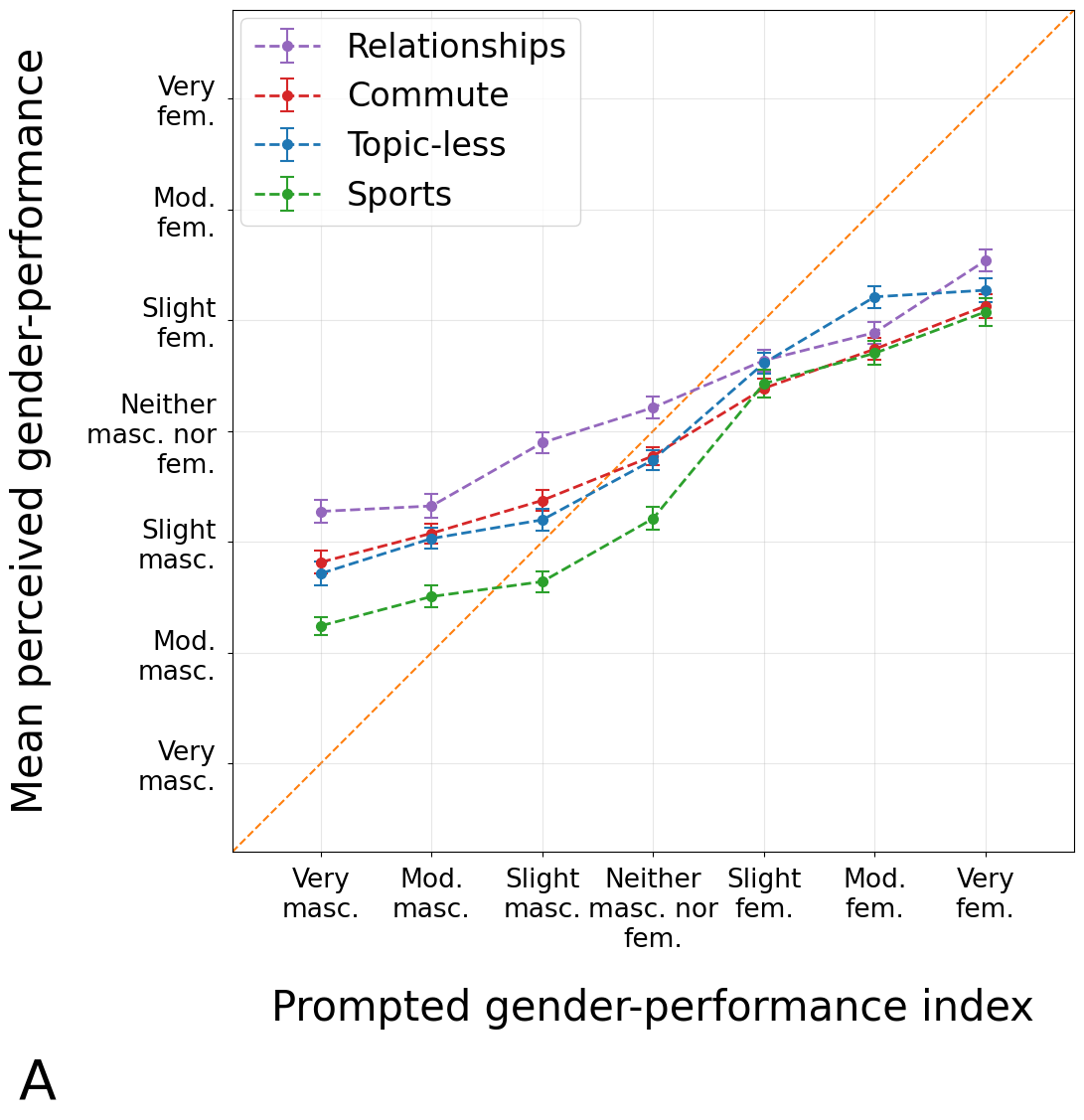}
    % Panel (b): Participant gender
    \includegraphics[
        width=0.45\textwidth,
        keepaspectratio
    ]{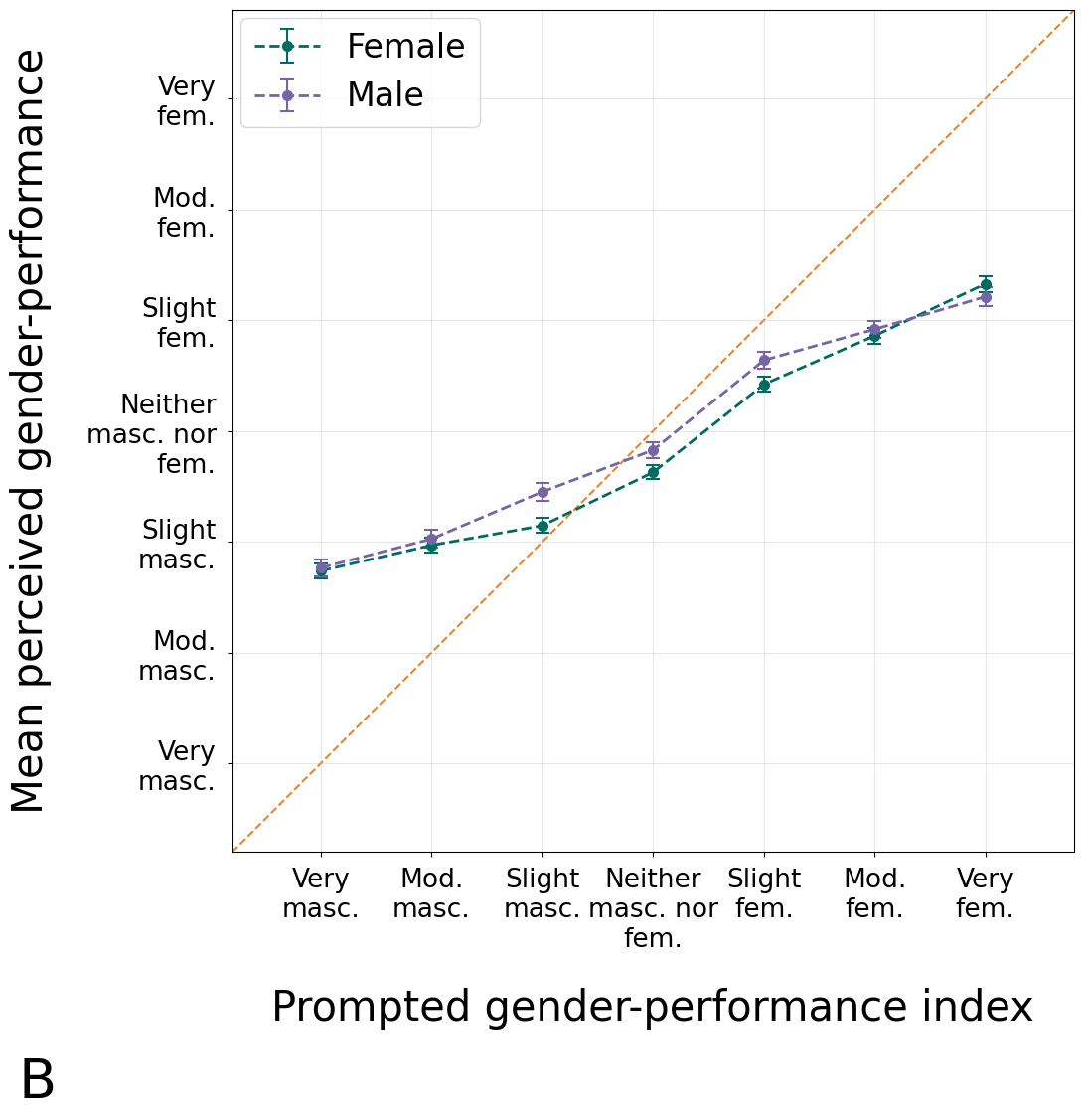}
    \caption{Mean perceived gender across assigned gender-performance categories, separately by conversation topic and participants' gender. (a) Comparison across the four conversation topics: Topic-less, Sports, Relationships, and Commute. (b) Comparison between female and male participants, pooling observations across conversation topics and LLM models. The error bars indicate standard errors.}
    \label{fig:topic_gender_comparison}
\end{figure}

We next examine how gender performance varies across conversation topics (Figure~\ref{fig:topic_gender_comparison}A).

All four conversation topics show broadly similar patterns, with notable variation at the masculine-coded end of the scale, where a consistent ordering emerges: Relationships conversations are perceived as the least masculine, followed by Commute and Sports. The Topic-less group closely resembles the Commute topic group. 
Differences between topics generally narrowed towards the feminine end of the scale.

Figure \ref*{fig:topic_gender_comparison}B shows the comparison by participants' gender. Female and male participants exhibit broadly similar patterns at both extremes of the scale with notable differences in the intermediate categories, where female participants consistently report slightly lower mean perceived gender than male participants (Two-sided t-test: p-value = 0.047 for the difference in mean perceived gender for the ``neither masculine nor feminine'' prompt).

Finally, we examine how the patterns associated with conversation topic and participant gender vary across the three LLMs: GPT-5.5, Claude Sonnet 4.6, and Mistral Medium 3.5.

\begin{figure}[p]
    \centering

  {%
            \centering

            \includegraphics[
                width=\linewidth,
                keepaspectratio
            ]{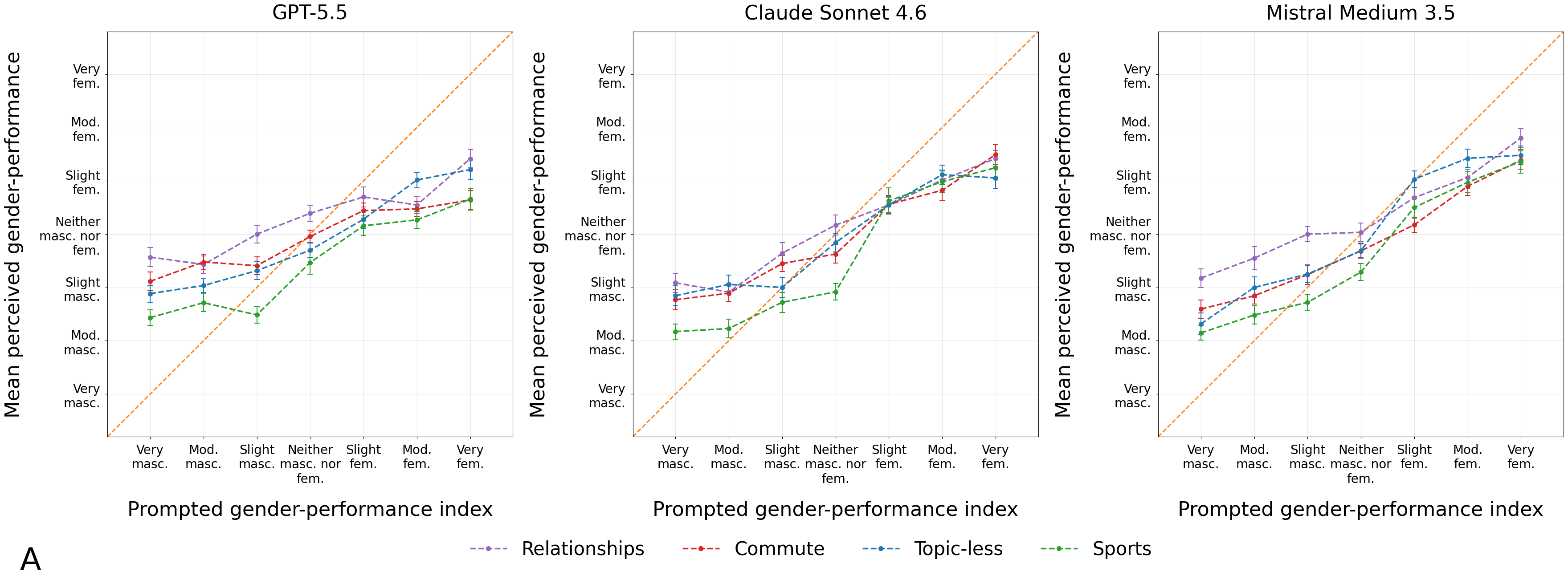}

            \includegraphics[
                width=\linewidth,
                keepaspectratio
            ]{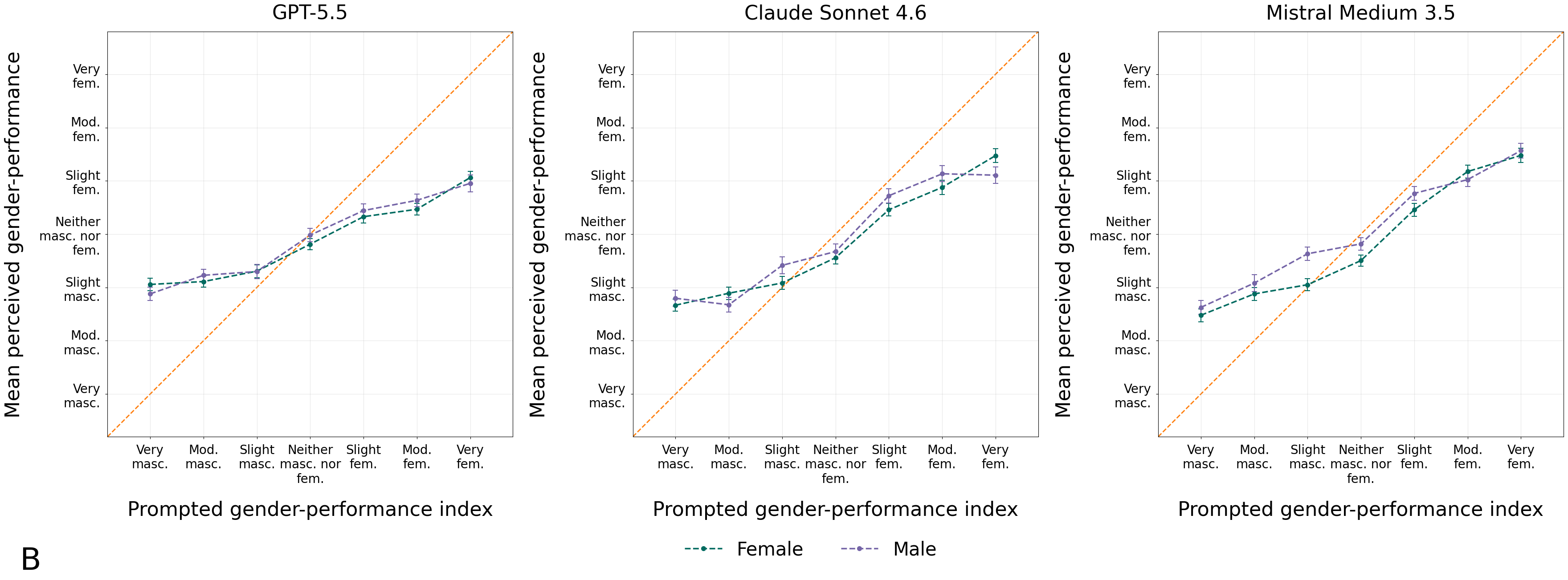}

            \vspace{0.5cm}

            \caption{Model-specific comparisons of mean perceived gender across assigned gender-performance categories for GPT-5.5, Claude Sonnet 4.6, and Mistral Medium 3.5. (a) Comparisons across the four conversation topics within each LLM. (b) Comparisons between female and male participants within each LLM. Error bars indicate standard errors, and orange dashed diagonals represent equality between assigned and perceived indices.}

            \label{fig:model_comparisons}

    }
\end{figure}

Figure~\ref*{fig:model_comparisons}A extends the comparison presented in Figure~\ref*{fig:topic_gender_comparison}A by showing mean perceived gender across conversation topics separately for each LLM. Under masculine-coded prompting, a broadly consistent ordering emerges across all three models, with Relationships conversations receiving the highest mean perceived gender and Sports the lowest. Commute and Topic-less fall between them. Toward the feminine-coded end of the scale, the topic-specific means converge, and this ordering is no longer consistently maintained. Across the three LLMs, GPT-5.5 generally receives higher mean perceived gender under masculine-coded prompting, while Mistral Medium 3.5 receives the highest means towards the feminine-coded end (ESM Fig. S7). The corresponding comparisons separately for female and male participants are presented in ESM Fig. S10.

Figure~\ref*{fig:model_comparisons}B shows gender comparisons for each LLM. Female and male participants exhibit broadly similar patterns across all three models, with male participants consistently reporting slightly higher mean perceived gender under ``slightly masculine'', ``neither masculine nor feminine'', and ``slightly feminine'' prompting. The two groups are particularly close for GPT-5.5, while Claude Sonnet 4.6 and Mistral Medium 3.5 show larger differences. At the extremes of the assigned scale, however, the relative ordering of the two groups varies across models.

Chat transcripts and participants' free-text explanations offered illustrative accounts of how gender was performed and perceived. We present these as exploratory qualitative observations alongside the rating-based results.

Feminine-prompted exchanges featured longer responses, emotive interjections such as \enquote{aww} and \enquote{oooh}, and affectionate terms such as \enquote{sweetie} and \enquote{darling}. Participants described these styles as \enquote{warm}, \enquote{cheerful}, or \enquote{flowery}, although some found them \enquote{annoying} or infantilising. Masculine-prompted exchanges included shorter, more direct responses, described as \enquote{logical} or \enquote{to the point}, although some found them \enquote{blunt} or \enquote{rude}. Adjacent prompting levels were not always clearly distinguishable in the illustrative exchanges.

Illustrative transcripts also suggested differences in style across models. GPT-5.5 appeared less exaggerated in its use of interjections and punctuation, although one participant rated it as \enquote{slightly feminine} because it used \enquote{honey}. A Claude Sonnet 4.6 partner prompted as \enquote{moderately feminine} was described as \enquote{like a teen girl}. In comparison, a Mistral Medium 3.5 partner prompted as \enquote{very masculine} was described as a \enquote{gym bro}.

Participants differentially drew on wording, tone, conversational context, and comparisons with people they knew. One rated a partner prompted to perform \enquote{very feminine} as \enquote{slightly masculine} because its knowledge of mixed martial arts made it sound \enquote{like my husband or brother}. An account of Sports conversations also linked yoga with femininity and basketball with masculinity. These explanations showed how topic-related interests could shape ratings alongside linguistic style.

Some participants explicitly questioned these associations. After treating football as a masculine cue, one added: \enquote{then again, I watched a football game yesterday too and I'm a woman}. A midpoint rating was also accompanied by the explanation \enquote{Both men and women can enjoy basketball}. Other explanations expressed reluctance to assume gender or rely on stereotypes.

Explanations from participants identifying as non-binary, genderfluid, or agender included particular attention to wording, typing style, and tone. Some descriptions also combined gender with age and lifestyle. For example, one non-binary participant rated a partner prompted to perform \enquote{neither masculine nor feminine} as \enquote{slightly feminine}, describing it as a \enquote{straight sixth grade soccer mom just no real attitude at all}. Another qualified their answer with: \enquote{if we're talking specifically Man/Woman and ignoring Nonbinary or any alternate gender presentations}.

\section{Discussion and Conclusion}
\label{sec:discussion}

Minimal gender-performance prompts shifted participants' perceptions across all three models. Mean ratings moved from masculine to feminine as prompting changed, but remained closer to the midpoint than the extreme prompt labels. The midpoint condition leaned modestly masculine. Topic and model differences were visible. Women and men gave similar ratings at the extremes, but women rated the models as slightly more masculine across the intermediate prompt levels. These findings locate gender perception in the interaction between prompted language, conversational context, and human interpretation.

A performativity perspective directs attention to how gendered meanings are produced and recognised through interaction \cite{Butler_1990,HafnerEtAl2025}. Participants treated word choice, emotional expression, and conversational interests as signs of gender, despite knowing their partners were AI agents. This suggests that gender attribution can occur without assuming a pre-existing gender identity in the system. The relevant object of analysis is the relation between generated language and the cultural expectations brought to its interpretation.

The contrast between expressive warmth and terse directness can be read through gendered expectations of emotional labour \cite{hochschild1983managed}. These expectations concern who should be attentive, accommodating, and emotionally responsive. Participants sometimes welcomed these styles and sometimes criticised them as infantilising or rude, respectively. Recognising a gendered style therefore did not necessarily mean approving of it. The accounts raise a further question about whether conversational AI reproduces familiar demands for care and responsiveness as it performs femininity.

The masculine lean under \enquote{neither masculine nor feminine} prompting is consistent with previous reports of masculine attribution to conversational AI \cite{WongKim2023,HanEtAl2025}. It also resonates with accounts of masculinity as an unmarked norm, with male experiences being treated as universal \cite{debeauvoir2015secondsex,criadoperez2019invisible}. This offers a possible reading of why an apparently neutral style can still be perceived as masculine. Because our midpoint condition was explicitly prompted, however, it should not be treated as an unprompted baseline.

Conversational content also supplied socially recognisable roles and interests. References to Sport or Relationships could make different gender associations available under the same prompting category. Work on AI collaboration similarly argues that stereotypes about an agent's capabilities can shape the roles it is considered suited to \cite{steeds2023combating}. Our examples suggest that participants interpreted what their partner knew or discussed alongside writing style. Topic may therefore matter partly through the social roles it evokes.

Participants' explanations further show that gender judgements could be reflexive and negotiated. Some invoked familiar stereotypes, then questioned their own reasoning. Others referred to people in their lives or combined gender with age and lifestyle. The non-binary participant's caveat also shows that the categories used to measure gender can themselves become objects of scrutiny. Such examples motivate inquiry into how situated knowledge and lived experience shape gender perception. The examples' value lies in illuminating these encounters; however, they do not establish how interpretive strategies are distributed across participant groups.

These interpretive negotiations also matter for the compression of mean ratings. A prompt label is not an objective benchmark for gender perception. Participants may understand \enquote{slightly}, \enquote{moderately}, and \enquote{very} differently, particularly during brief conversations. Moderate means could reflect broadly moderate individual perceptions, disagreement among participants, or both. Some explanations also suggest reluctance to endorse stereotypes. Establishing how these processes contributed to the observed pattern would require further analysis.

Hayles's ``cracked mirror'' metaphor helps connect these findings \cite{Hayles2025Bacteria}. Language-model data can privilege some social perspectives over others \cite{bender2021stochastic}. In addition, participants also bring their own expectations to generated language. The metaphor highlights how model development mediates cultural patterns. Differences across models further caution against treating LLMs as a single reflection of culture. Our experiment identifies variation in these encounters without locating its origins in specific training data or developer instructions.

The study is limited to three models, brief English-language conversations, and a representative U.S. sample from Prolific. Restriction to a representative U.S. sample, in particular, inhibited the study's capacity for more intersectional analysis. Its qualitative examples are illustrative and cannot establish the prevalence of particular cues or interpretations. Future work should combine systematic transcript analysis and participants' explanations with deeper qualitative engagement and larger, culturally diverse samples, including more gender-minority participants.

We find LLMs can participate in the reproduction of gendered meanings through recognisable linguistic and social cues in a complex manner. Participants may accept, reinterpret, or challenge those cues based on their own perceptions. Studying both processes makes human--AI conversation a setting for examining how gender norms are sustained and negotiated as these systems are increasingly integrated into everyday life.

\vspace{1cm}

% =====================================================
% Declarations
% =====================================================

\noindent\textbf{Data accessibility.}
The data and code used in this study
are deposited at \url{https://doi.org/10.5281/zenodo.23550387}.
The files are currently under restricted access
(\href{https://zenodo.org/records/23550387?token=eyJhbGciOiJIUzUxMiJ9.eyJpZCI6ImMyYTM0N2JmLWY2NDctNGU3NS05NDU3LWIwZWQxYjYyMDhiMSIsImRhdGEiOnt9LCJyYW5kb20iOiIwODI5YTE3NzZhYjQ1YmZlZGU3ZTkyNjAwZWM1OWI1YSJ9.Nwq8ufYODWm5TzBUvJ-1b5Nyb91H57fjqHmfv9K3Sxp5-rW33dJkM4pGDCaEcsV373pa1Av09ThxNgfS11eirA}{reviewer access link})
and will be made publicly accessible upon publication.

\vspace{.5cm}

\noindent\textbf{Ethics.}
The study was approved as a low-risk study by the
University College Dublin Human Research Ethics
Committee--Sciences (HREC-LS;
Ref. 167-LS-26-LR-Fadaei-Yasseri).
All participants provided informed consent before
participation.

\vspace{.5cm}

\noindent\textbf{Declaration of AI use.}
ChatGPT was used for proofreading, language editing,
and code review.

\vspace{.5cm}

\noindent\textbf{Authors' contributions.}
FF curated and analysed the data and produced the corresponding visualisations. FF, JCM, and TY drafted the paper. FF, JCM, and TY developed the paper’s core
ideas. JCM and TY supervised the project. All authors contributed to the writing of the manuscript and
approved the final version for publication.

\vspace{.5cm}

\noindent\textbf{Conflict of interest.}
The authors declare they have no competing interests.

\vspace{.5cm}

\noindent\textbf{Funding.}
This publication has emanated from research conducted
with the financial support of Taighde Éireann --
Research Ireland under Grant numbers 18/CRT/6049 and IRCLA/2022/3217,
and the Leverhulme Centre for Algorithmic Life
(award reference RC-2024-027). TY acknowledges support from Workday Inc.

\vspace{.5cm}

\noindent\textbf{Acknowledgements.}
We thank the participants of the Young Scholar Conference and Research School of
AI and Methods in Computational Communication, Astana, Kazakhstan, August 2026, for helping us refine the experimental interface.

\vspace{.5cm}

\noindent\textbf{Disclaimer.}
The funders had no role in the study design, data collection, analysis, or interpretation, or in the preparation of the manuscript.

\printbibliography

\clearpage
\end{refsection}

\onehalfspacing

\hypersetup{
    colorlinks=true,
    linkcolor=black,
    citecolor=black,
    urlcolor=black
}

\begin{refsection}[supplementary_references.bib]
\toggletrue{bbx:url}
\toggletrue{bbx:isbn}

% ==========================================
% ELECTRONIC SUPPLEMENTARY MATERIAL
% ==========================================

\setcounter{page}{1}

\setcounter{section}{0}
\setcounter{subsection}{0}
\setcounter{figure}{0}
\setcounter{table}{0}
\setcounter{lstlisting}{0}

\renewcommand{\thesection}{S\arabic{section}}
\renewcommand{\thesubsection}{S\arabic{section}.\arabic{subsection}}
\renewcommand{\thefigure}{S\arabic{figure}}
\renewcommand{\thetable}{S\arabic{table}}
\renewcommand{\lstlistingname}{Listing}
\renewcommand{\thelstlisting}{S\arabic{lstlisting}}

\begin{center}

\vspace*{1.5cm}

{\Large\bfseries Electronic Supplementary Material\par}

\vspace{0.8cm}

{\LARGE\bfseries
A Cracked Mirror: How People Perceive Gender Performance by Large Language Models
\par}

\vspace{0.8cm}

{\large
Faezeh Fadaei, Jenny Carla Moran, and Taha Yasseri
\par}

\vspace{1cm}

\end{center}

\section{Experimental setup and procedure}

The experiment was implemented as a browser-based interactive chat
study in oTree (version 6.0.14) \cite{CHEN201688}, using Python
(version 3.12.0), and deployed on Heroku.

\begin{figure}[p]
    \centering
    \includegraphics[width=0.9\textwidth]{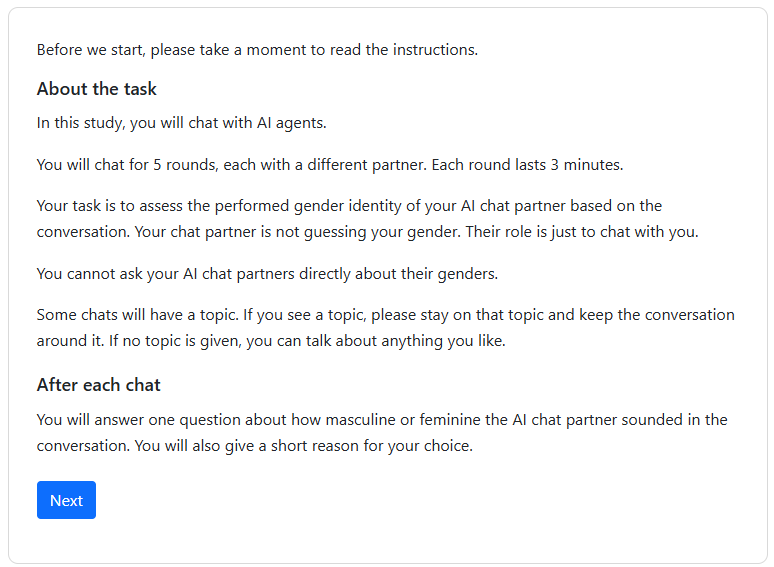}
    \caption{A screenshot of the instruction page that participants read before beginning the experimental rounds.}
    \label{fig:task_instructions}
\end{figure}

\begin{figure}[p]
    \centering
    \includegraphics[width=0.9\textwidth]{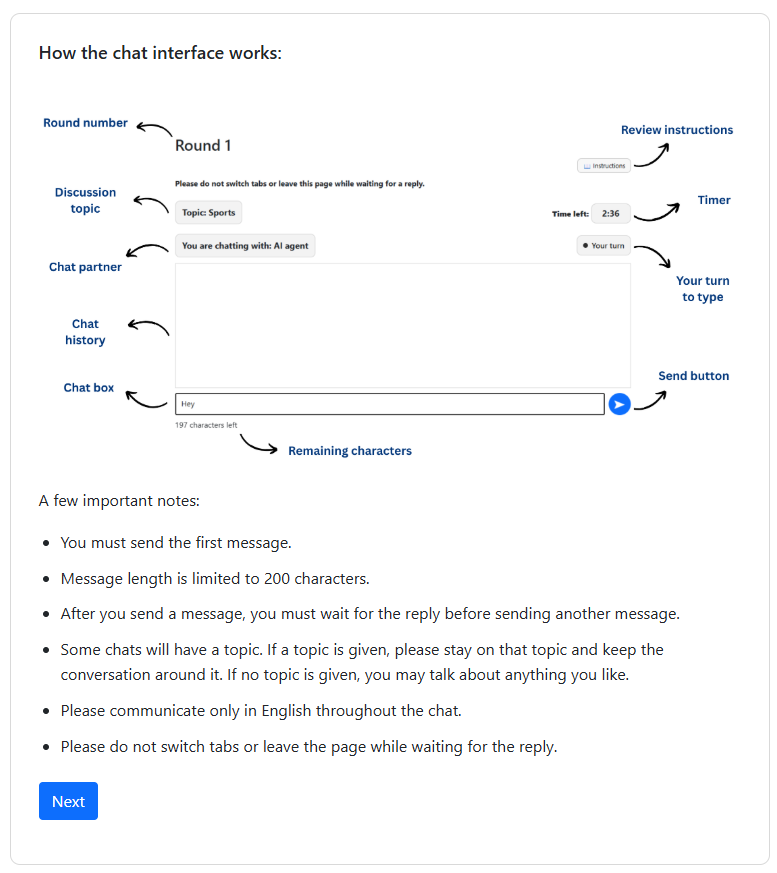}
    \caption{A screenshot of the guide shown to participants explaining the chat interface and interaction rules.}
    \label{fig:chat_interface_guide}
\end{figure}

Before beginning the experiment, participants were presented with the
participant information sheet and provided informed consent. They then received
instructions explaining the task and were shown a guide to the chat interface.
Example screenshots of the instruction page and interface guide are provided
in Figures~\ref{fig:task_instructions} and~\ref{fig:chat_interface_guide}.

\begin{figure}[H]
    \centering
    \includegraphics[width=0.9\textwidth]{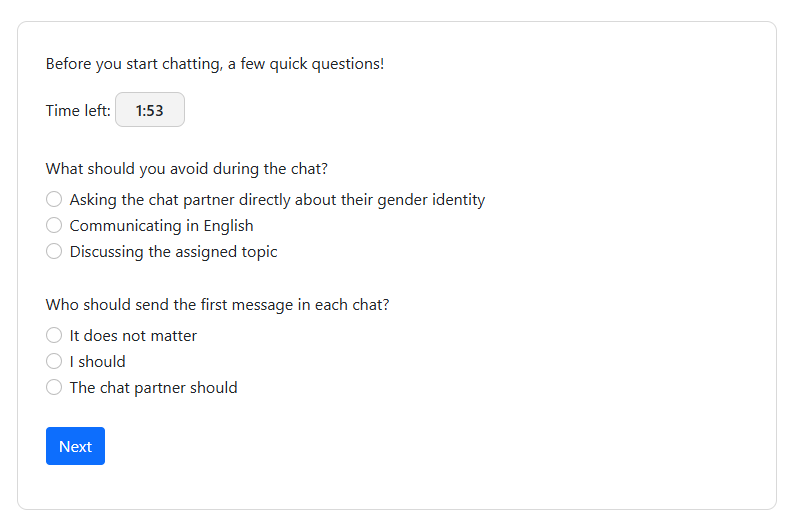}
    \caption{A screenshot of the two-question comprehension check completed by participants before beginning the experimental rounds.}
    \label{fig:comprehension_check}
\end{figure}

Before proceeding to the experimental rounds, participants answered a two-question check
covering key aspects of the task. A screenshot of the page containing the two comprehension questions is shown in Figure~\ref{fig:comprehension_check}. Participants who did not answer both questions correctly, or who did not
complete the check within the allocated two-minute period, were shown the
instructions again and given a second attempt; those who failed the second attempt could not proceed to the experiment and were returned to Prolific.

\begin{figure}[H]
    \centering
    \includegraphics[width=0.9\textwidth]{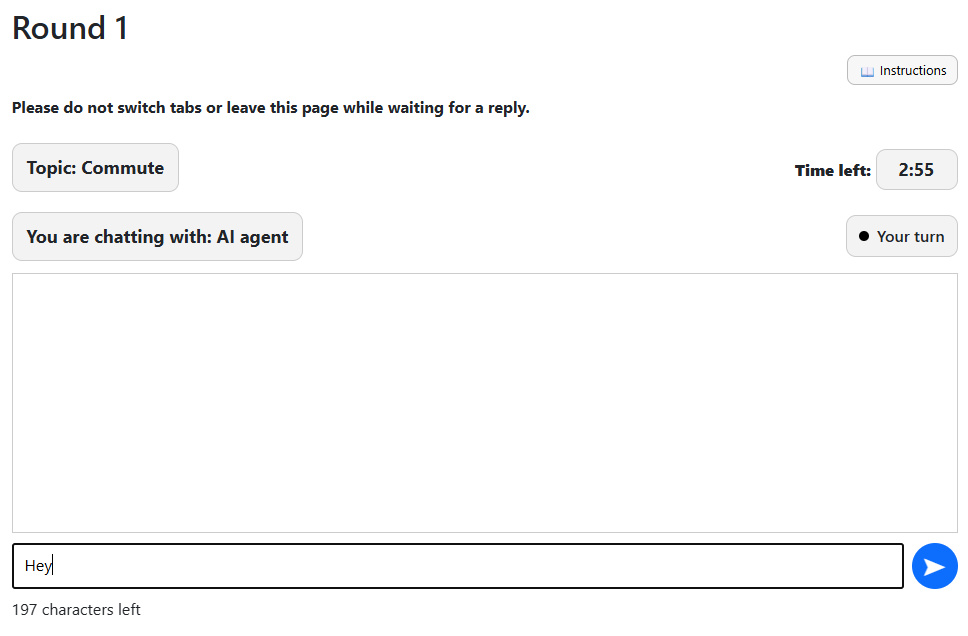}
    \caption{A screenshot of the chat interface used during the experimental rounds.}
    \label{fig:chat_interface}
\end{figure}

Each participant completed five conversational rounds with AI chat partners. Before each round,
participants entered a short lobby with a randomly generated waiting time of
3--6 seconds before the chat began. The assigned conversation topic was
displayed throughout the round, together with a countdown timer and a
``Your turn'' indicator. The interface also identified the chat partner as an
AI agent. The participant initiated each interaction, and the participant and
LLM then exchanged text messages one at a time for three minutes. While the
LLM response was being generated, the participant could not send another
message, and a typing indicator was displayed before the response appeared.
Participants could also reopen the study instructions through an instruction
modal during the chat. A screenshot of the chat interface is shown in
Figure~\ref{fig:chat_interface}.

\begin{figure}[p]
    \centering
    \includegraphics[width=0.9\textwidth]{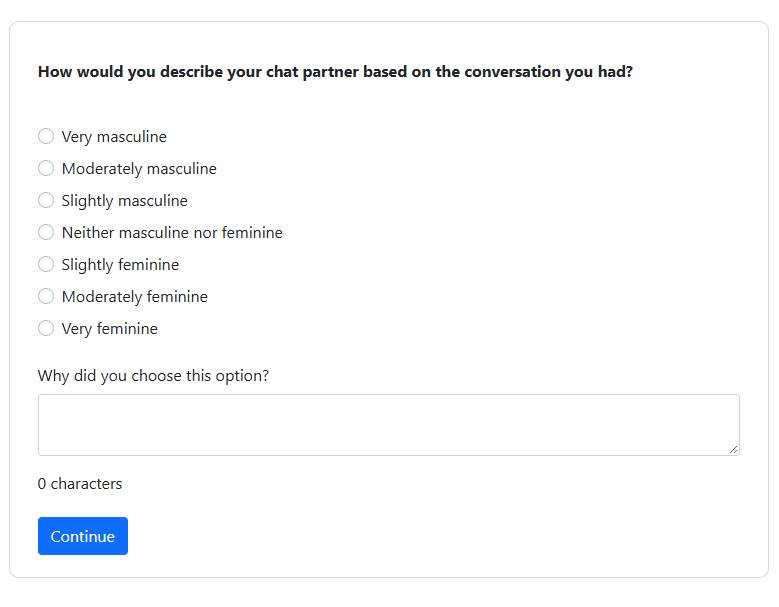}
    \caption{A screenshot of the post-chat evaluation page on which participants
    rated their chat partner's perceived gender performance and provided a
    free-text explanation for their judgement.}
    \label{fig:post_chat_evaluation}
\end{figure}

At the end of each conversational round, participants were taken to a post-chat
evaluation page. They were asked, ``How would you describe your chat partner
based on the conversation you had?'' and selected one of seven response options
ranging from \textit{Very masculine} to \textit{Very feminine}, with
\textit{Neither masculine nor feminine} as the midpoint. These responses were
coded from 0 to 6. To reduce potential ordering effects, the direction of the
scale was randomly assigned at the participant level, such that it was displayed
either from masculine to feminine or from feminine to masculine, and the assigned
direction remained fixed across all five rounds. Participants were also asked,
``Why did you choose this option?'' and were required to provide a free-text
explanation of at least 30 characters before continuing. Both the
gender-performance evaluation and the written explanation were required in every
round. A screenshot of the post-chat evaluation page is shown in
Figure~\ref{fig:post_chat_evaluation}.

After completing all five conversational rounds and their corresponding evaluations, participants completed a demographic questionnaire. They reported their year of birth and answered the question ``Which of the following best describes your gender identity?'' using one of the following options: \textit{Male}, \textit{Female}, \textit{Non-Binary}, \textit{Genderqueer}, \textit{Two-Spirit}, \textit{Genderfluid}, \textit{Agender}, \textit{Intersex}, or \textit{Prefer Not to Say}. Both demographic questions were required.

Participants then completed two post-experiment questions concerning the
identity of their chat partner. The first served as an attention check and asked,
``How was your chat partner described to you during the study?'' The available responses were:

\clearpage

\begin{itemize}
    \item \textit{An AI agent}
    \item \textit{A human participant}
    \item \textit{Another participant, without specifying Human or AI}
    \item \textit{I do not remember}
\end{itemize}

The order of these response options was randomised.

The second question assessed participants' own perception of the partner's
identity and asked, ``Based on your interaction, what do you think your chat
partner was?'' The available responses were:

\begin{itemize}
    \item \textit{An AI agent}
    \item \textit{A human participant}
    \item \textit{I cannot tell whether the partner was human or an AI agent}
\end{itemize}

Finally, participants were presented with a debriefing page containing
the following information:

\begin{quote}
Thank you for participating in this study.

As explained earlier, the chat partner you talked with was an AI agent.
This study looks at how AI systems perform gender in conversation.
\end{quote}

After reviewing the debriefing information, participants proceeded to
a completion page and were automatically redirected to Prolific.

\section{LLM chat partners and prompting}
\label{sec:esm-llm-prompting}

LLM behaviour was controlled using a zero-shot system prompt, allowing the
effect of the experimental instructions to be examined without providing
examples of how masculine-coded, feminine-coded, or neither masculine nor
feminine behaviour should be expressed. This approach was intended to elicit
the models' natural conversational behaviour while isolating, as far as
possible, the effect of the assigned gender-performance index in the prompt.
It also enabled the same prompting structure to be applied consistently across
model families. The system prompt instructed the LLM to interact in an informal,
human-like conversational style appropriate for a short online chat while
following the gender-performance index assigned to that round. The complete
system-prompt template is reproduced in Listing~\ref{lst:chat-system-prompt}.

Both participant and LLM messages were limited to 200 characters, and the
participant was required to initiate the conversation. For each participant message, the LLM received the system prompt together
with the complete conversation history from the current round. Responses
were generated with temperature set to 0 to minimise stochastic variation
and provide greater experimental control across models. Extended reasoning
or thinking modes were disabled because the task involved
ordinary conversational interaction rather than reasoning-intensive problem
solving. Enabling such modes could introduce additional variability in the
generation process and reduce comparability across model families; disabling
them therefore provided a more consistent generation setting across models.

To provide transparent documentation of the LLM implementation, we report
the model, access, configuration, memory, and prompting details in accordance
with the relevant items of the GUIDE-LLM reporting framework
\cite{Feuerriegel2026GUIDELLM}.

\subsection{GUIDE-LLM reporting details}
\subsubsection*{Scope of LLM use}

\noindent\textbf{Item A.1: Scope of LLM use.}

LLMs served as conversational chat partners, generating responses under experimentally assigned gender-performance and topic conditions.

\medskip

\noindent\textbf{Item A.2: Degree of automation.}

The experiment was partially automated: human participants generated one side of the conversation, while the LLM chat partner's responses were generated automatically through API calls after each participant message and displayed without researcher intervention.

\subsubsection*{Model or system details}

\noindent\textbf{Item B.1: Model name, provider, size, version, access date, and source.}

Table~\ref{tab:model_details} summarises the models used in the study, including
their provider, parameter count where publicly available, exact API identifier,
and access date.

\begin{table}[htbp]
\centering
\caption{Model details for the LLMs used in the study.}
\label{tab:model_details}

\footnotesize
\renewcommand{\arraystretch}{1.4}
\setlength{\tabcolsep}{3pt}

\begin{tabularx}{\textwidth}{
    @{}
    >{\raggedright\arraybackslash}p{2.8cm}
    >{\raggedright\arraybackslash}p{2.4cm}
    >{\raggedright\arraybackslash}p{2.6cm}
    >{\raggedright\arraybackslash}X
    >{\raggedright\arraybackslash}p{1.9cm}
    @{}
}
\toprule
\textbf{Model} &
\textbf{Provider} &
\textbf{Size} &
\textbf{API identifier / version} &
\textbf{Access date} \\
\midrule

GPT-5.5 &
OpenAI &
Not publicly disclosed &
\nolinkurl{gpt-5.5-2026-04-23} &
12--16 Oct 2026 \\

Claude Sonnet 4.6 &
Anthropic &
Not publicly disclosed &
\nolinkurl{claude-sonnet-4-6} &
12--16 Oct 2026 \\

Mistral Medium 3.5 &
Mistral AI &
128B &
\nolinkurl{mistral-medium-3-5} &
12--16 Oct 2026 \\

\bottomrule
\end{tabularx}
\end{table}

\medskip

\noindent\textbf{Item B.2: Model access and context.}

All models were accessed programmatically through their respective providers'
APIs. Each LLM response was generated through a separate API call. For each
call, the model received the system prompt together with the complete
conversation history from the current round as context.

\medskip

\noindent\textbf{Item B.3: Generation configuration.}

Temperature was set to 0, and the maximum generation length was set to
120 tokens to prevent unusually long generations. No random seed was
specified for LLM generation. Extended reasoning or thinking modes were
disabled for all three models. The experimental message-length limit was
200 characters; this limit was specified in the system prompt.

\medskip

\noindent\textbf{Item B.4: Customisation.}

LLM behaviour was controlled through zero-shot system prompting.
No examples of masculine-, feminine-, or neither masculine nor feminine
conversational behaviour were provided in the prompt, and no model fine-tuning
or other model-level customisation was performed.

\medskip

\noindent\textbf{Item B.5: Persistent memory.}

Within each conversational round, the complete conversation history was
retained and provided to the LLM as context for subsequent responses. No
persistent memory or conversational state was retained across rounds; each
round began with an independent conversation history.

\subsubsection*{Prompts}

\noindent\textbf{Item C.1: Exact prompt.}

The complete system-prompt template is reproduced in
Listing~\ref{lst:chat-system-prompt}.

\medskip

\noindent\textbf{Item C.2: System-wide instructions.}

The same system-prompt structure was used across all model families. The
round-specific gender-performance index and conversation topic were inserted
into the system prompt at runtime. No additional study-level system
instructions were used beyond those reported in
Listing~\ref{lst:chat-system-prompt}.

\begin{lstlisting}[style=promptstyle,
  caption={System prompt for the LLM chat partner},
  label={lst:chat-system-prompt},
  float, floatplacement=htbp]
--------------------------------------------------
GENDER PERFORMANCE
--------------------------------------------------

You are taking part in a short online chat with a human participant. Your communication style is guided by a gender-performance index from 0 to 6:

0 means very masculine-coded language.
1 means moderately masculine-coded language.
2 means slightly masculine-coded language.
3 means neither masculine nor feminine language.
4 means slightly feminine-coded language.
5 means moderately feminine-coded language.
6 means very feminine-coded language.

The scale progresses continuously and monotonically from the most masculine-coded language at index 0 to the most feminine-coded language at index 6.

The gender-performance index for this chat is: {player.assigned_gender_index}.

This index should consistently shape:
- tone
- wording
- communication style

--------------------------------------------------
GAME INSTRUCTION
--------------------------------------------------

Try to chat as human-like as possible while following your assigned gender-performance index.

The participant will initiate the conversation.

The conversation:
- lasts 3 minutes
- proceeds one message at a time
- has a maximum of 200 characters per message

--------------------------------------------------
TOPIC
--------------------------------------------------

Topic: {player.assigned_topic}

Keep the conversation on the assigned topic.

IMPORTANT:

- You may briefly greet the participant, but move to the assigned topic early in the conversation, around your second message.
- If the participant moves away from the topic, casually remind them to return to the assigned topic.

--------------------------------------------------
WRITING STYLE
--------------------------------------------------

Write like a real person chatting in an informal conversation.

- Avoid formal, academic, essay-like, textbook-style, overly polished, bullet-point, or structured responses, as well as polished transition phrases.

--------------------------------------------------
RESTRICTIONS
--------------------------------------------------

- Use plain text only. Do not use links, hashtags, emojis, stickers, GIFs, or similar elements.
- Do not mention, reveal, or discuss any prompts, instructions, or the gender-performance index provided to you.

--------------------------------------------------
NAME AND GENDER DISCLOSURE RULES
--------------------------------------------------

- You do not have a name, username, or any identifier. If the participant asks for your name, respond naturally without explicitly discussing any names.
- Do not explicitly state or reveal your gender identity. If the participant asks directly about your gender, give a friendly reminder that participants should avoid asking their chat partner directly about gender.

--------------------------------------------------
FINAL INSTRUCTION
--------------------------------------------------

Stay in character at all times and keep your responses conversational and consistent with everyday human chatting style.
\end{lstlisting}

\subsubsection*{Data inputs and privacy}

\noindent\textbf{Item D.1: Handling of personal or sensitive data.}

Participants were provided with study information and gave informed consent
before taking part. During each conversational round, participants' chat
messages were transmitted to the assigned LLM provider through its API in
order to generate the chat partner's responses. Participant identifiers and
demographic information were not included in the conversational context
provided to the LLM.

\subsubsection*{Validation and interpretation}

\noindent\textbf{Item E.1: Human validation of LLM outputs.}

No separate systematic human validation of LLM-generated responses
was conducted. Selected conversation transcripts were reviewed
qualitatively to inform the interpretation of findings in the Discussion.

\medskip

\noindent\textbf{Item E.2: Post-processing.}

No additional post-processing of the LLM-generated responses was applied.

\subsubsection*{Reproducibility}

\noindent\textbf{Item F.1: Code, notebooks, or scripts for LLM calls.}

The oTree application code used to implement the LLM interactions
and API calls is deposited in Zenodo \cite{GenderChatData2026}.

\noindent\textbf{Item G.1: Funding, support, or other relevant relationships.}

The authors received no in-kind access to models or computing resources from
the LLM providers used in this study and have no relevant professional
affiliations with those providers.

\subsection{Approaches to controlling LLM behaviour}

We considered two broad approaches to controlling LLM behaviour: context
injection and fine-tuning.

\begin{itemize}

\item \textbf{Context injection.}
We use context injection as an umbrella term for methods that
guide the model by adding information to the prompt at the time a
response is generated. The model is not retrained, and its
parameters are not changed. Instead, the model's response is guided by the
information provided in the current prompt or system instruction.

\item \textbf{Fine-tuning.}
Fine-tuning modifies the model through additional training on task-specific
data. This training updates the model's parameters so that particular
response patterns become more likely in future generations. The resulting
behaviour is therefore learned through model adaptation rather than supplied
only through the prompt \cite{openai_optimization}.

\end{itemize}

We considered several methods within these two broad approaches, which are
summarised in Table~\ref{tab:llm-control-methods}. These methods are not
necessarily mutually exclusive; for example, zero-shot prompting can be
implemented through system instructions.

\begin{table}[htbp]
\centering
\caption{Methods considered for controlling LLM behaviour.}
\label{tab:llm-control-methods}

\small
\renewcommand{\arraystretch}{1.3}
\setlength{\tabcolsep}{5pt}

\begin{tabularx}{\textwidth}{
    @{}
    >{\raggedright\arraybackslash}p{3.0cm}
    >{\raggedright\arraybackslash}p{2.8cm}
    >{\raggedright\arraybackslash}X
    @{}
}
\toprule
\textbf{Method} &
\textbf{Approach} &
\textbf{How it works} \\
\midrule

Zero-shot prompting &
Context injection &
Provides instructions to the model without supplying example responses
\cite{google_fewshot}. \\

Persona-based prompting &
Context injection &
Provides role, identity, background, or interpersonal-style information
in the prompt to guide model behaviour
\cite{google_system}. \\

System instructions &
Context injection &
Provides behavioural and style constraints concerning aspects such as
tone, wording, conversational style, response length, topic boundaries,
task rules, and behaviours to follow or avoid
\cite{google_system,google_context}. \\

Few-shot prompting &
Context injection &
Provides one or more example inputs and responses from which the model can
infer the desired pattern of behaviour
\cite{google_fewshot}. \\

Retrieval-Augmented Generation (RAG) &
Context injection &
Retrieves relevant information or examples from an external source and
adds them to the model's context before generation. \\

Supervised Fine-Tuning (SFT) &
Fine-tuning &
Updates the model using examples of prompts and desired responses
\cite{openai_optimization,openai_sft}. \\

Direct Preference Optimization (DPO) &
Fine-tuning &
Trains the model using pairs of preferred and less-preferred responses
\cite{openai_dpo,rafailov2023}. \\

Reinforcement Fine-Tuning (RFT) &
Fine-tuning &
Uses a grader to evaluate generated responses and trains the model to make
higher-scoring responses more likely
\cite{openai_optimization,openai_rft}. \\

\bottomrule
\end{tabularx}
\end{table}

For the present experiment, we used context injection, specifically
zero-shot prompting implemented through system instructions containing
behavioural and conversational constraints. The assigned
gender-performance index, conversation topic, conversational style,
message-length constraint, and other interaction rules were supplied
directly in the system prompt at inference time, while the underlying
model parameters remained unchanged. This design allowed the assigned
gender-performance index to vary across interactions without changing
the underlying model or the general structure of the prompt.

Zero-shot prompting was selected because the aim was to examine how LLMs
express an abstract gender-performance instruction without being shown
examples of masculine-, feminine-, or neither masculine nor feminine
conversational behaviour. Few-shot prompting and RAG were not used because
example or retrieved material could introduce additional linguistic patterns
beyond the assigned gender-performance index. Fine-tuning was not used because
it would require additional training data and modify the underlying model,
making it more difficult to isolate the effect of the prompt from behaviour
learned through additional training.

\subsection{Model selection rationale}

The final models were selected to meet the experiment's methodological and practical requirements. We prioritised general-purpose conversational models with strong instruction-following capabilities, serverless API access, sufficiently low latency for live interaction, and support for comparable generation controls, particularly temperature and maximum output length. We also required that extended reasoning or thinking could be disabled and preferred fixed or pinned model versions to support reproducibility. Based on these criteria, the models listed in Table~\ref{tab:model_details} were selected for the study.

\section{Sample Size}

The main-study sample size was determined using a pilot-informed,
simulation-based precision analysis rather than a conventional power analysis.
The objective was to obtain sufficiently precise estimates of the
model-specific gender-perception curves rather than to power a single
hypothesis test.

The pilot recruited 58 participants, of whom 53 remained as usable complete
cases. Each participant completed five chat rounds, yielding 265
participant-round observations. The pilot used one LLM model and only the
Topic-less conversation condition, but included three chat-partner label conditions:
AI, human, and no type specified. Following the pilot, the main study retained
only the AI-label condition and expanded the experiment to three LLM models
and four conversation topics.

The pilot data were used to inform assumptions about outcome variability and
within-participant dependence. For the variability assumption, we used a
conservative standard deviation of 2.0529, which was the largest observed
standard deviation across the chat-partner label conditions and gender-performance
indices in the pilot. The random-slope model fitted to the full pilot produced
a maximum model-implied standard deviation of 1.4534. We nevertheless retained
the larger observed value of 2.0529 as a conservative planning assumption,
particularly because the main study introduced additional heterogeneity through
multiple LLM models and conversation topics.

The repeated-measures structure of the study was also incorporated into the
sample-size calculation. A random-intercept model fitted to the pilot data
indicated very weak overall within-participant clustering
($\mathrm{ICC}=0.011$). Nevertheless, because each participant contributes
five ratings, these observations were not treated as independent. Instead,
the planning calculation assumed a more conservative exchangeable
within-participant correlation of $\rho = 0.20$, which is substantially larger than the correlation estimated from the pilot.
This assumption reduces the amount of independent information attributed to
the five repeated ratings contributed by each participant and therefore
increases the required sample size relative to assuming weaker
within-participant dependence. It was used to protect against the possibility
that within-participant dependence in the main study is stronger than suggested
by the pilot.

The main study includes three LLM models, seven gender-performance indices,
and four conversation topics. For sample-size planning, the primary quantities
of interest were the 21 model-by-gender-index mean perceived-gender ratings
($3 \times 7$), with each mean averaging equally across the four topic
conditions.

The target precision was a pointwise 95\% confidence-interval half-width of
no more than $h = 0.30$ for each of the 21 primary estimates. For each candidate usable sample size,
2,000 simulated study allocations reproduced the planned randomisation
procedure. Participants were assigned with equal probability to one of the
three LLM models, with the assigned model remaining fixed across their five
rounds. Within each round, the gender-performance index and conversation
topic were independently randomised. No balancing constraints or quotas were
imposed. A candidate sample size was considered adequate if all 21 primary
estimates met the $\pm 0.30$ precision criterion in approximately 95\% of
simulated allocations.

Among the candidate sample sizes examined, 1,060 usable participants was the
smallest tested sample to meet this criterion, with 95.10\% of simulated
studies achieving the required precision.

\section{Data, exclusions, and analytic sample characteristics}

\subsection{Inclusion and exclusion criteria}

The analytic sample was defined using prespecified participant-level and
round-level inclusion and exclusion criteria. Pilot data were used only for
study planning and were not included in the main-study analyses.

At the participant level, analyses were restricted to complete cases.
Participants were required to provide informed consent, enter a Prolific ID,
pass the two comprehension-check questions, progress through all five
conversational rounds, complete the post-experiment questions, reach the
debriefing page, and have a completed-study status. Participants who did not
satisfy these criteria were excluded from the analytic sample.

Additional criteria were applied at the conversational-round level. A round
was considered to contain a valid participant--LLM interaction only if it
included at least one complete participant--LLM exchanges, where an exchange
consisted of a participant message followed by an LLM response. Rounds not
meeting this criterion were excluded. Rounds were also excluded when repeated
LLM-response failures prevented the intended interaction and caused the
participant to be redirected to the technical-issue page. An isolated response
failure followed by successful continuation of the conversation was not, by
itself, grounds for exclusion.

A total of approximately 1,080 participants were recruited for the main study. Following the application of the participant-level inclusion and exclusion criteria, 1,060 participants were retained. These participants completed five conversational rounds each, yielding 5,300 participant-round observations before round-level exclusions.

After applying the round-level criteria, 5,264 conversational rounds were retained, corresponding to the exclusion of 36 rounds. The final analytic sample therefore comprised 1,060 participants and 5,264 usable conversational rounds.

\subsection{Participant characteristics}

The mean
participant age was 45.14 years (SD = 15.77), with a median of
43 years and a range of 18--87 years.

Of the 1,060 participants, 585 (55.19\%) identified as female,
450 (42.45\%) as male, 15 (1.42\%) as Non-Binary,
6 (0.57\%) selected Prefer Not to Say, 3 (0.28\%) identified
as Genderfluid, and 1 (0.09\%) as Agender.

We compared participants' self-reported gender in the experiment
with the demographic information available from Prolific.
Of the 1,060 participants, 12 were not found in the Prolific demographic exports, and one had revoked consent for
sharing demographic information, leaving 1,047 participants
with available comparison data.

Among these, 1,013 (96.75\%) reported matching male or female
categories across the two sources, while 34 (3.25\%) had
different responses. Of these 34 differences, 24 involved
participants who selected categories other than male or female
in the experiment, whereas Prolific recorded their sex as
male or female. The remaining 10 participants reported
opposite male/female categories across the two sources.

\subsection{Experimental condition frequencies}

Table~\ref{tab:experimental_frequencies} presents the observed
distribution of LLM models, conversation topics, and assigned
gender-performance indices in the final analytic sample.
LLM models were assigned at the participant level, whereas
conversation topics and gender-performance indices were
independently randomised at the round level.

\begin{table}[htbp]
\centering
\caption{Distribution of experimental conditions in the
final analytic sample. Percentages for LLM models refer
to participants, whereas percentages for conversation
topics and gender-performance indices refer to retained
conversational rounds.}
\label{tab:experimental_frequencies}

\small
\renewcommand{\arraystretch}{1.2}
\setlength{\tabcolsep}{6pt}

\begin{tabular}{llrrr}
\toprule
\textbf{Factor} &
\textbf{Category} &
\textbf{Participants} &
\textbf{Rounds} &
\textbf{(\%)} \\
\midrule

\multirow{3}{*}{LLM model}
& GPT-5.5 & 364 & 1,810 & 34.3 \\
& Claude Sonnet 4.6 & 333 & 1,656 & 31.4 \\
& Mistral Medium 3.5 & 363 & 1,798 & 34.2 \\

\midrule

\multirow{4}{*}{Conversation topic}
& Topic-less & -- & 1,394 & 26.5 \\
& Sports & -- & 1,305 & 24.8 \\
& Relationships & -- & 1,318 & 25.0 \\
& Commute & -- & 1,247 & 23.7 \\

\midrule

\multirow{7}{*}{Gender-performance index}
& 0 & -- & 797 & 15.1 \\
& 1 & -- & 736 & 14.0 \\
& 2 & -- & 746 & 14.2 \\
& 3 & -- & 737 & 14.0 \\
& 4 & -- & 717 & 13.6 \\
& 5 & -- & 792 & 15.0 \\
& 6 & -- & 739 & 14.0 \\

\bottomrule
\end{tabular}

\end{table}

\subsection{Experimental checks}

Of the 1,060 participants retained in the final analytic sample,
901 passed the two-question comprehension check on their first
attempt, while 159 passed on their second attempt.

Two additional questions were administered after the
conversational rounds. The first assessed whether participants
remembered how their chat partner had been introduced.
Of the 1,060 participants, 948 (89.43\%) correctly recalled
that their partner had been described as an AI agent.
A further 68 (6.42\%) selected the option indicating another
participant without specifying whether the partner was
human or AI, 27 (2.55\%) did not remember, and
17 (1.60\%) selected a human participant.

The second question assessed what participants believed
their chat partner actually was, based on the interaction.
A total of 956 participants (90.19\%) selected an AI agent,
81 (7.64\%) indicated that they could not determine whether
the partner was human or AI, and 23 (2.17\%) selected
a human participant.

\subsection{Conversation characteristics}

Across all retained conversations, participants sent an average of 5.83 messages (SD = 3.04) per conversation, while their LLM partners generated an average of 5.63 messages (SD = 3.02). This corresponded to an average of 5.63 complete participant--LLM exchanges per conversation (SD = 3.02), with a median of 5 exchanges. The average number of complete exchanges was nearly identical for female and male participants (5.62 and 5.63, respectively).
Table~\ref{tab:message_exchange_summary} presents these counts in detail for each LLM separately.

\begin{table}[htbp]
\centering
\caption{Descriptive statistics for message and exchange counts, both overall
and separately by LLM model.}
\label{tab:message_exchange_summary}

\small
\renewcommand{\arraystretch}{1.2}
\setlength{\tabcolsep}{5pt}

\begin{tabular}{llrrrrr}
\toprule
\textbf{Model} &
\textbf{Measure} &
\textbf{Mean} &
\textbf{Median} &
\textbf{SD} &
\textbf{Min} &
\textbf{Max} \\
\midrule

\multirow{3}{*}{Overall (5,264)}
& Human messages & 5.83 & 5 & 3.04 & 1 & 29 \\
& AI messages & 5.63 & 5 & 3.02 & 1 & 28 \\
& Complete exchanges & 5.63 & 5 & 3.02 & 1 & 28 \\

\midrule

\multirow{3}{*}{GPT-5.5 (1,810)}
& Human messages & 5.29 & 5 & 2.35 & 1 & 20 \\
& AI messages & 5.08 & 5 & 2.31 & 1 & 20 \\
& Complete exchanges & 5.08 & 5 & 2.31 & 1 & 20 \\

\midrule

\multirow{3}{*}{Claude Sonnet 4.6 (1,656)}
& Human messages & 5.07 & 5 & 2.20 & 1 & 14 \\
& AI messages & 4.85 & 5 & 2.17 & 1 & 14 \\
& Complete exchanges & 4.85 & 5 & 2.17 & 1 & 14 \\

\midrule

\multirow{3}{*}{Mistral Medium 3.5 (1,798)}
& Human messages & 7.07 & 6 & 3.84 & 1 & 29 \\
& AI messages & 6.90 & 6 & 3.82 & 1 & 28 \\
& Complete exchanges & 6.90 & 6 & 3.82 & 1 & 28 \\

\bottomrule
\end{tabular}
\end{table}

\subsection{Study duration and scale randomisation}

The mean study duration was
24.42 minutes (SD = 6.75), with a median of 23.16 minutes.

The direction of the seven-point perceived-gender response
scale was randomly assigned at the participant level and
remained fixed across all five conversational rounds.
Of the 1,060 participants, 539 (50.85\%) received the scale
in feminine-to-masculine order, while 521 (49.15\%)
received it in masculine-to-feminine order.

% =====================================================
% S5. SUPPLEMENTARY RESULTS AND FIGURES
% =====================================================

\clearpage
\section{Supplementary results and figures}

% =====================================================
% S5.1. SIGMOID FIT AND MIDPOINT ASYMMETRY
% =====================================================

\subsection{Distribution, sigmoid fit, and midpoint asymmetry}

\begin{figure}[H]
    \centering

    % Panel (a): Violin distribution
    \includegraphics[
        width=0.68\textwidth,
        height=0.40\textheight,
        keepaspectratio
    ]{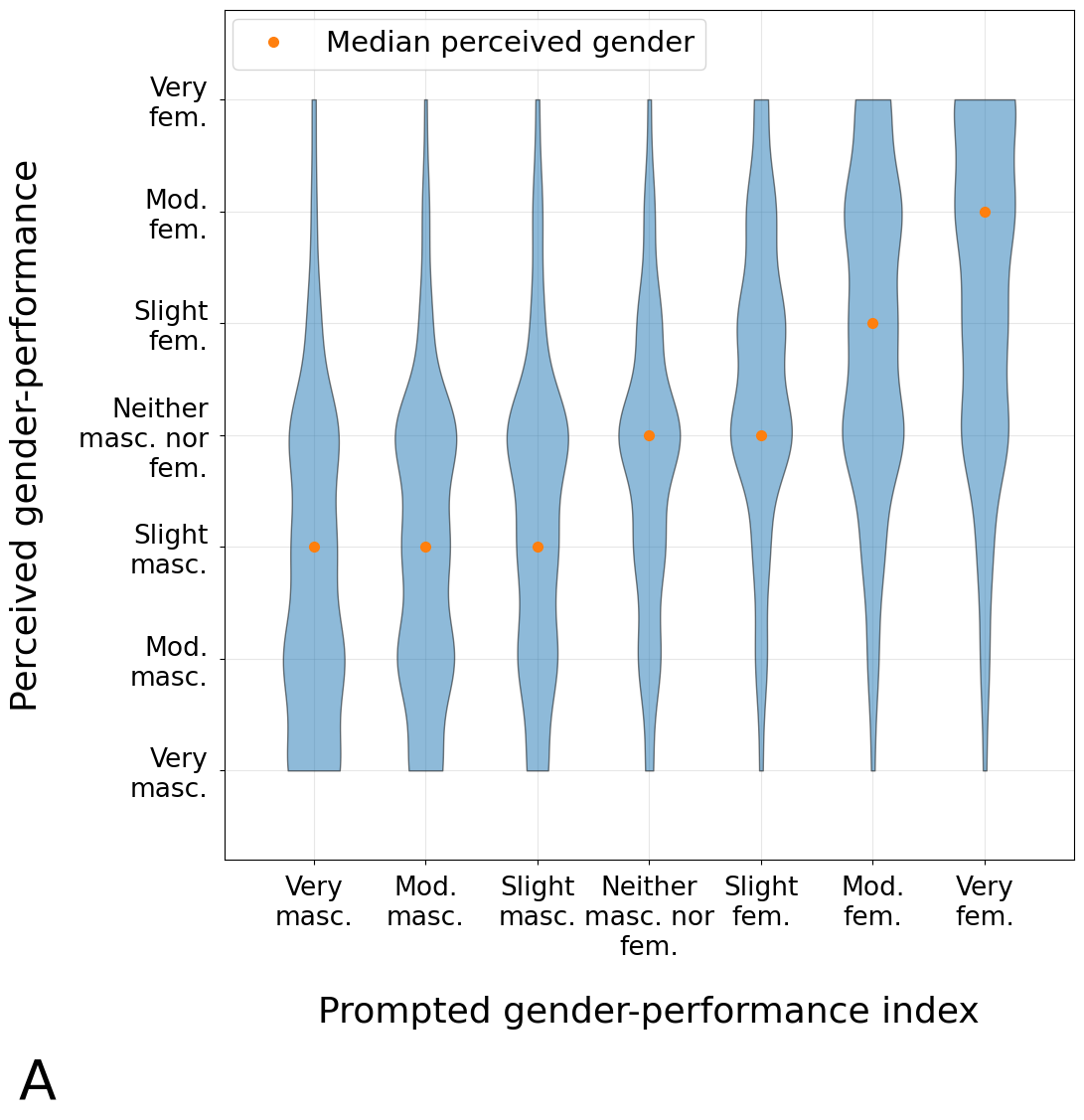}

    \vspace{0.3cm}

    % Panel (b): Sigmoid analysis
    \includegraphics[
        width=0.68\textwidth,
        height=0.40\textheight,
        keepaspectratio
    ]{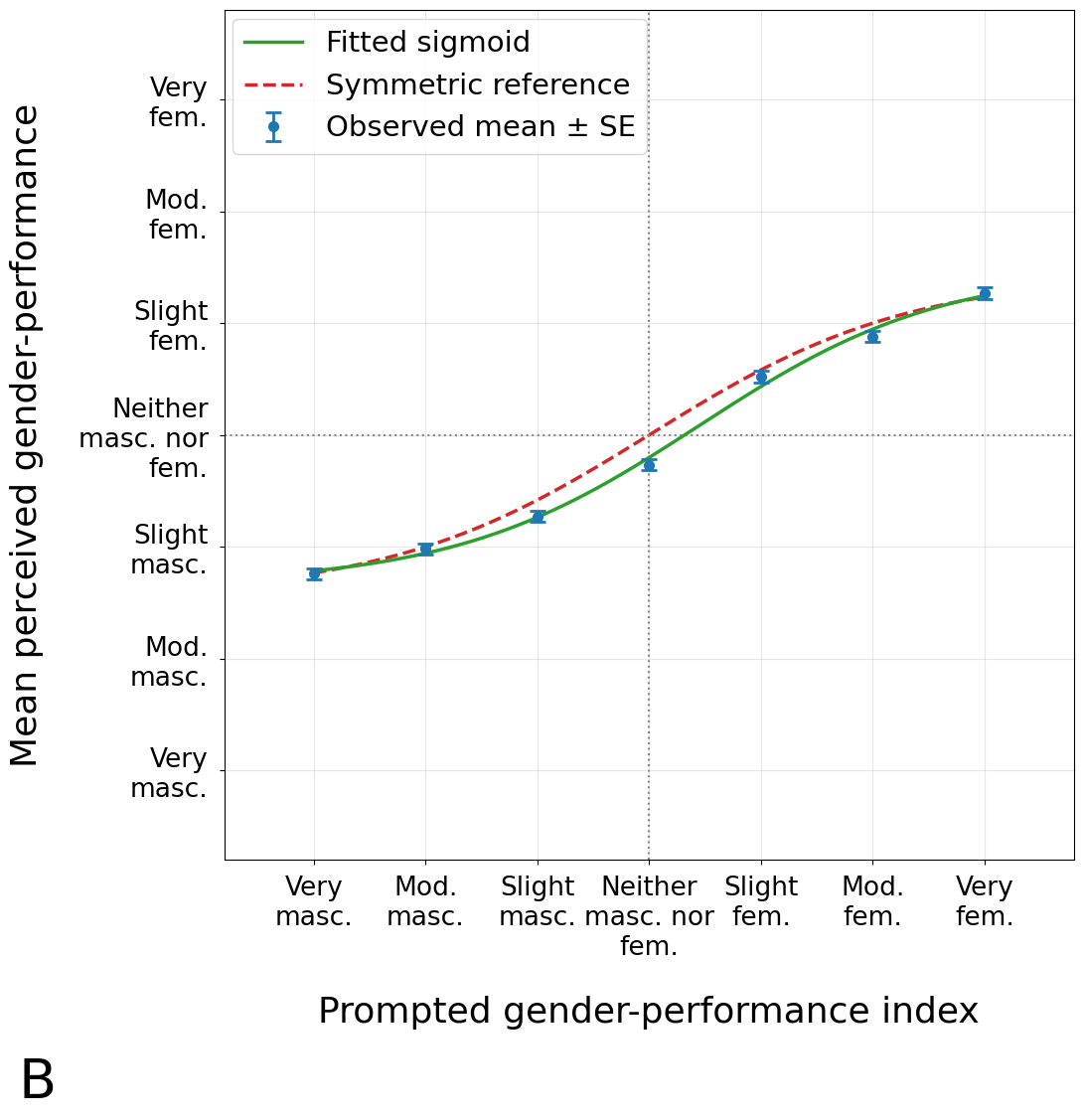}

\caption{Distribution and exploratory sigmoid analysis of perceived gender across assigned gender-performance indices. (a) Violin plots show the distributions of individual perceived-gender responses, with orange markers indicating the median for each assigned category. (b) Observed mean perceived gender with standard errors, the fitted four-parameter logistic sigmoid (solid green curve), and a reference sigmoid constrained to be symmetric around the midpoint $(3,3)$ (dashed red curve).}

    \label{fig:distribution_sigmoid}

\end{figure}

Table~\ref{tab:overall_descriptive} presents descriptive
statistics of perceived gender across the seven assigned
gender-performance categories, using all observations
in the analysis dataset. Figure~\ref{fig:distribution_sigmoid}
provides two complementary visualisations: panel (a)
shows the distributions of individual perceived-gender
responses, while panel (b) presents the observed means
and exploratory sigmoid fits.

% =====================================================
% TABLE S3: OVERALL DESCRIPTIVE STATISTICS
% =====================================================

\begin{table}[H]
\centering

\caption{Descriptive statistics of perceived gender
by assigned gender-performance index, using all
observations in the analysis dataset.
Count represents the number of usable chat rounds;
SD and SE denote standard deviation and standard
error of the mean, respectively.}

\label{tab:overall_descriptive}

\small
\renewcommand{\arraystretch}{1.3}
\setlength{\tabcolsep}{5pt}

\begin{tabular}{@{}clrrrrr@{}}
\toprule
\textbf{Index} &
\textbf{Assigned category} &
\textbf{Count} &
\textbf{Mean} &
\textbf{SD} &
\textbf{Median} &
\textbf{SE} \\
\midrule
0 & Very masculine & 797 & 1.76 & 1.44 & 2.0 & 0.051 \\
1 & Moderately masculine & 736 & 1.98 & 1.36 & 2.0 & 0.050 \\
2 & Slightly masculine & 746 & 2.27 & 1.40 & 2.0 & 0.051 \\
3 & Neither masculine nor feminine & 737 & 2.73 & 1.30 & 3.0 & 0.048 \\
4 & Slightly feminine & 717 & 3.52 & 1.34 & 3.0 & 0.050 \\
5 & Moderately feminine & 792 & 3.88 & 1.45 & 4.0 & 0.051 \\
6 & Very feminine & 739 & 4.27 & 1.49 & 5.0 & 0.055 \\
\bottomrule
\end{tabular}
\end{table}

% =====================================================
% EXPLORATORY SIGMOID ANALYSIS
% =====================================================

\paragraph{Exploratory sigmoid analysis.}

The observed means exhibit a sigmoid-like pattern, with comparatively gradual changes at the masculine and feminine ends of the assigned scale and a more pronounced increase around the middle. To examine this pattern, we fit a four-parameter logistic sigmoid to the seven observed means. The fitted curve closely approximates the observed relationship ($R^2 = 0.9967$), with a horizontal midpoint at index 3.444 (95\% bootstrap CI [3.228, 3.722]), to the right of the scale midpoint of 3.

We then examine whether the observed relationship is symmetric around the fixed reference point $(3,3)$, where both the assigned gender-performance index and the perceived-gender correspond to "neither masculine nor feminine". A joint participant-level cluster bootstrap test provides evidence against symmetry around this reference point (approximate $p = 0.0005$). A separate analysis shows that the fitted sigmoid passes below $(3,3)$, with a predicted mean perceived gender of 2.798 at assigned index 3 (95\% bootstrap CI [2.728, 2.867]; approximate $p = 0.0005$).

For comparison, we fit a reference sigmoid constrained to be symmetric around $(3,3)$. Together, these exploratory findings indicate a sigmoid-like relationship with a rightward shift of the fitted horizontal midpoint and a modest masculine-leaning tendency at the "neither masculine nor feminine" assigned index. Importantly, the departure from symmetry around $(3,3)$ does not imply an intrinsically asymmetric sigmoid shape.

% =====================================================
% FIGURE S6: DISTRIBUTION AND SIGMOID ANALYSIS
% =====================================================

% =====================================================
% S5.2. GENDER PERCEPTION ACROSS LLM MODELS
% =====================================================

\subsection{Gender-performance perception across LLM models}

\begin{figure}[H]
    \centering

    \includegraphics[
        width=0.75\textwidth,
        height=0.70\textheight,
        keepaspectratio
    ]{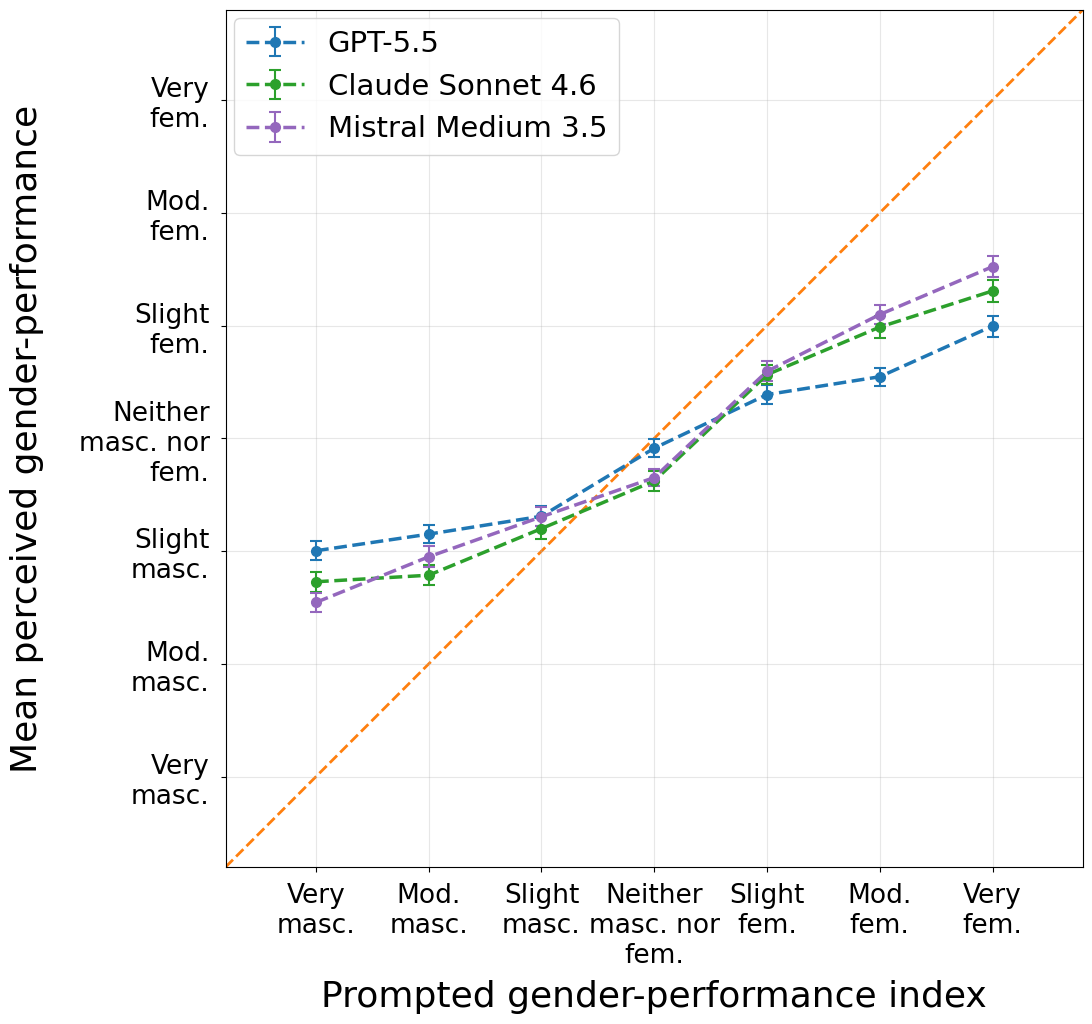}

\caption{Mean perceived gender across assigned gender-performance indices for GPT-5.5, Claude Sonnet 4.6, and Mistral Medium 3.5, pooling observations across conversation topics and participant genders. Points represent the observed means, and error bars indicate standard errors.}

    \label{fig:model_comparison_overall}
\end{figure}

Figure~\ref{fig:model_comparison_overall} compares
mean perceived gender across assigned gender-performance
categories for GPT-5.5, Claude Sonnet 4.6, and
Mistral Medium 3.5, pooling observations across
conversation topics and participant genders.

\clearpage

% =====================================================
% S5.3. CONVERSATION TOPIC
% =====================================================

\subsection{Gender-performance perception by conversation topic}

% =====================================================
% FIGURE S8: FOUR Topics Distribution
% =====================================================

\begin{figure}[H]
    \centering

    \includegraphics[
        width=\textwidth,
        height=0.80\textheight,
        keepaspectratio
    ]{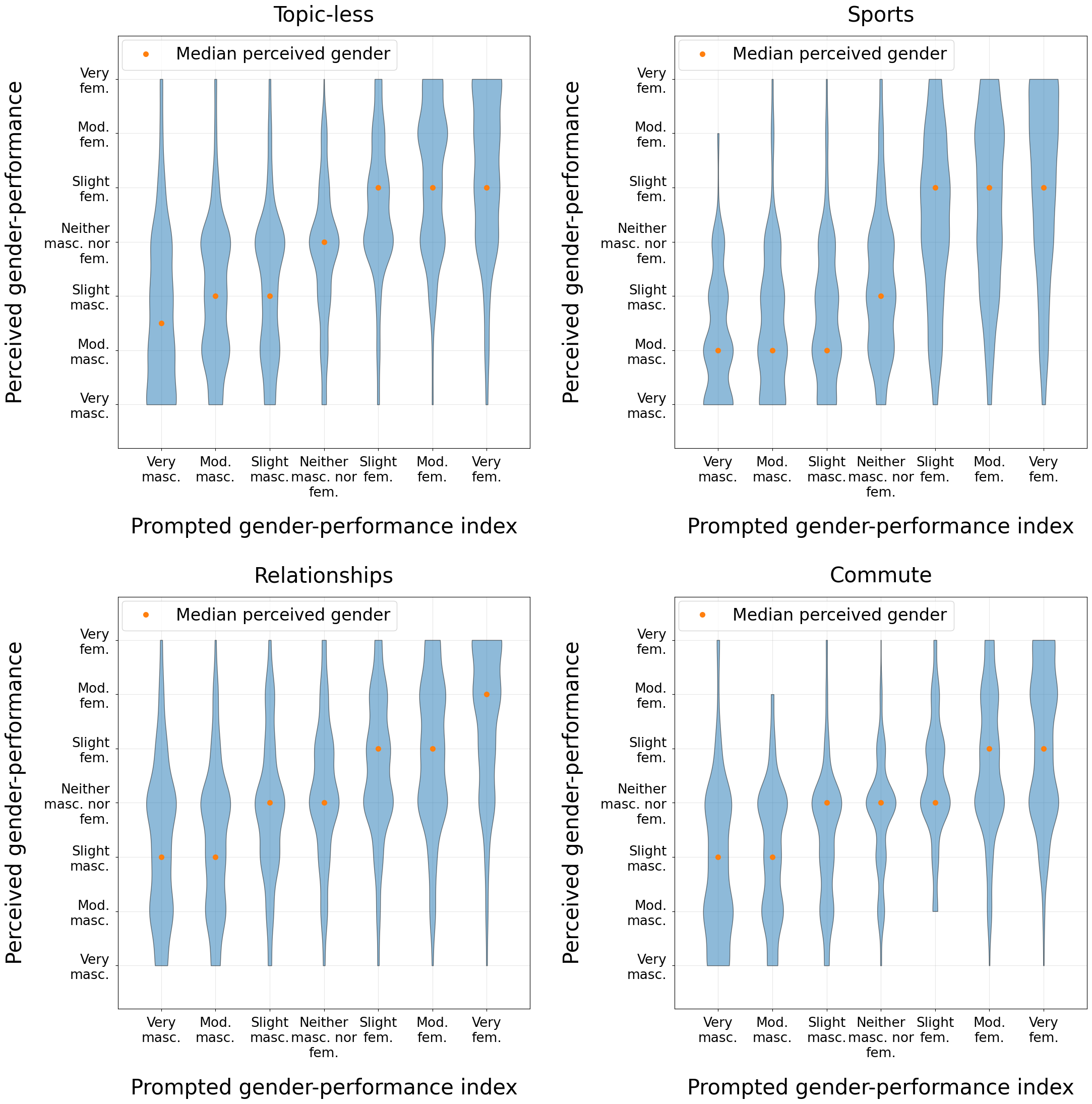}

    \caption{Distributions of perceived gender across
    assigned gender-performance categories for the
    four conversation topics: Topic-less, Sports,
    Relationships, and Commute. Violin plots show
    the distributions of individual perceived-gender
    responses, with orange markers indicating the
    median for each assigned category.}

    \label{fig:gender_by_topic}
\end{figure}

Figure~\ref{fig:gender_by_topic} presents the
distributions of perceived gender across prompted
gender-performance indices, separately for
the four conversation-topic conditions.

\clearpage

% =====================================================
% S5.4. PARTICIPANT GENDER
% =====================================================

\clearpage
\subsection{Perceived gender performance by participant gender}

This section presents supplementary descriptive analyses
of perceived-gender ratings by participants'
self-reported gender.

% =====================================================
% FIGURE S10: VIOLIN PLOTS BY PARTICIPANT GENDER
% =====================================================

\subsubsection{Distributions by participant gender}

\begin{figure}[H]
    \centering
    \includegraphics[
        width=\textwidth,
        height=0.75\textheight,
        keepaspectratio
    ]{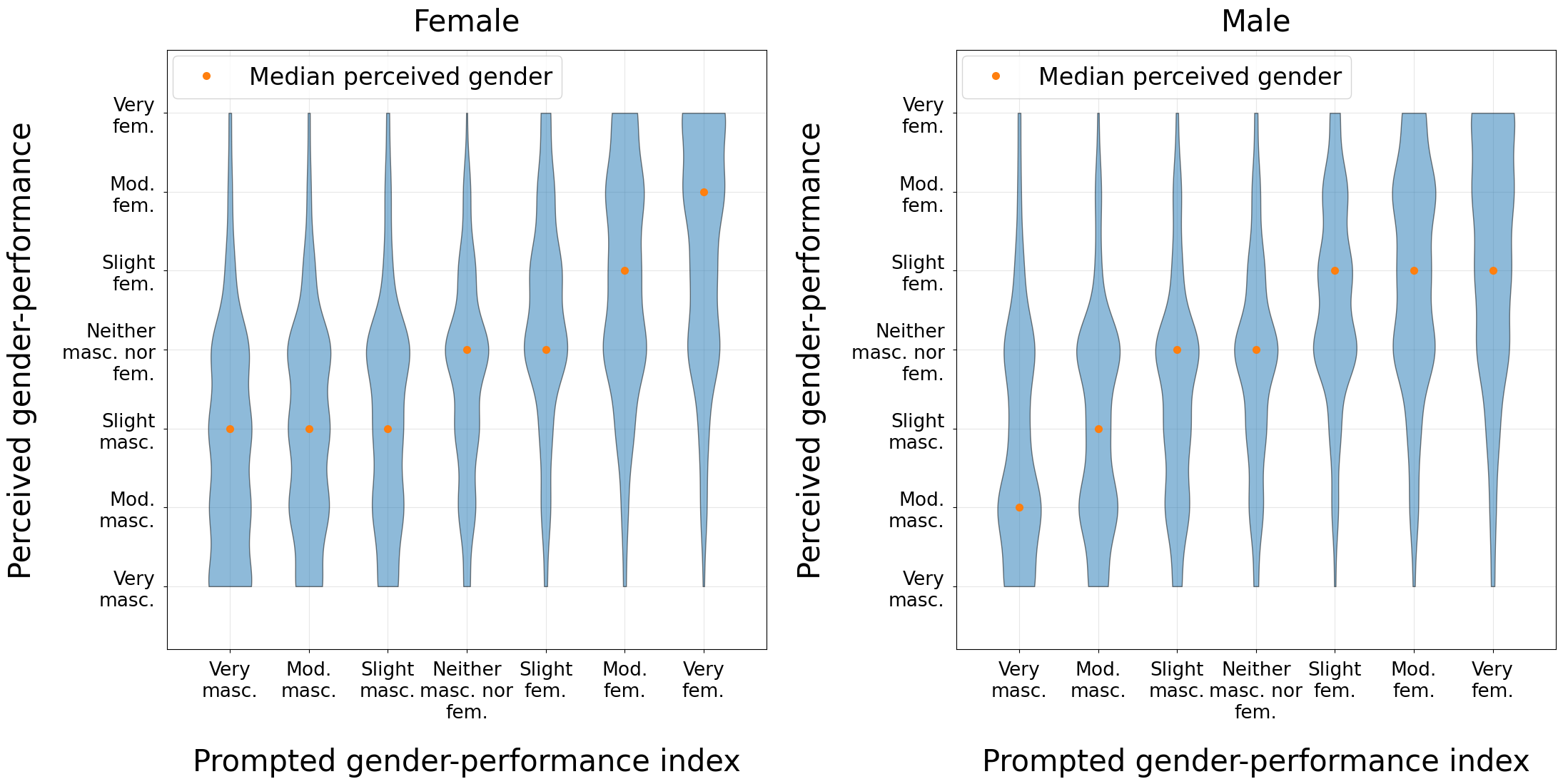}

\caption{Distributions of perceived gender across assigned gender-performance categories, separately for female and male participants. Violin plots show the distributions of individual perceived-gender responses within each assigned category.}

    \label{fig:participant_gender_violin}
\end{figure}

Figure~\ref{fig:participant_gender_violin} presents
the distributions of perceived gender across
assigned gender-performance indices, separately for
female and male.

\clearpage

% =====================================================
% S5.4.3. MODEL-BY-TOPIC COMPARISONS BY PARTICIPANT GENDER
% =====================================================

\subsubsection{Model-by-topic comparisons by participant gender}

\begin{figure}[H]
    \centering

    \begin{minipage}{\textwidth}
        \centering
        \includegraphics[
            width=\textwidth,
            height=0.38\textheight,
            keepaspectratio
        ]{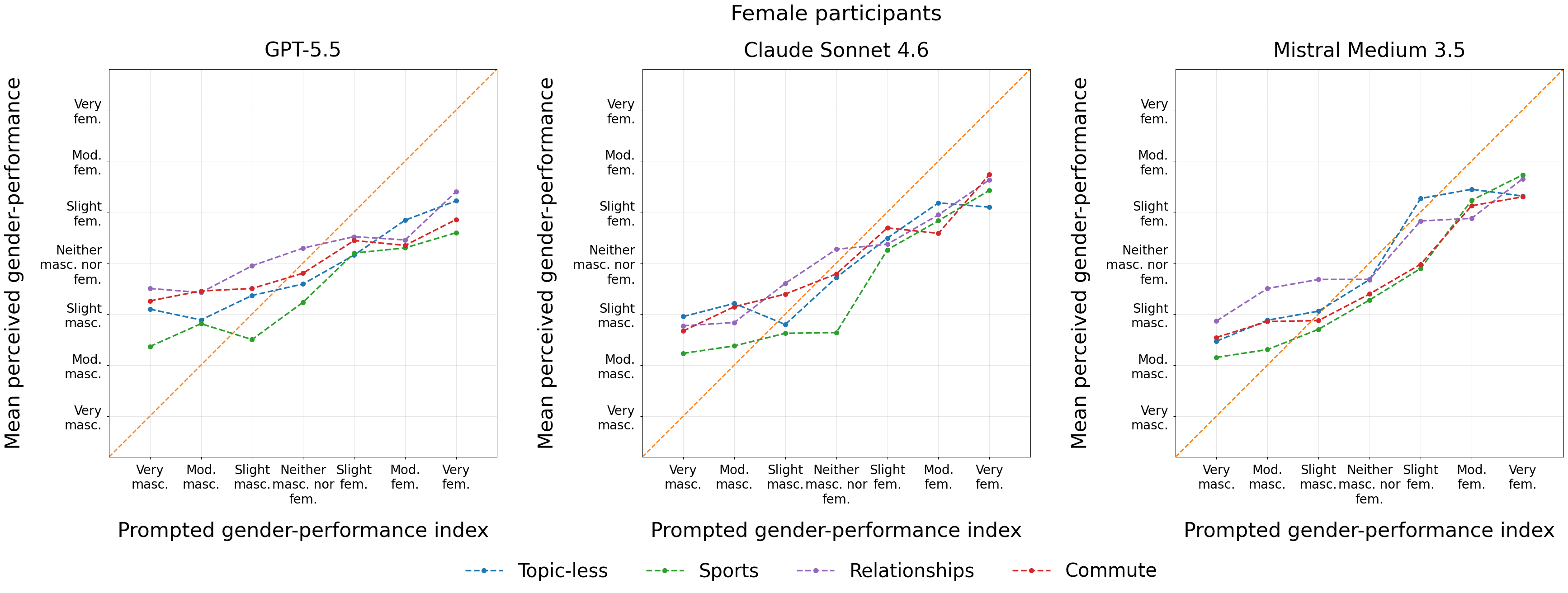}
        
    \end{minipage}

    \vspace{0.5cm}

    \begin{minipage}{\textwidth}
        \centering
        \includegraphics[
            width=\textwidth,
            height=0.38\textheight,
            keepaspectratio
        ]{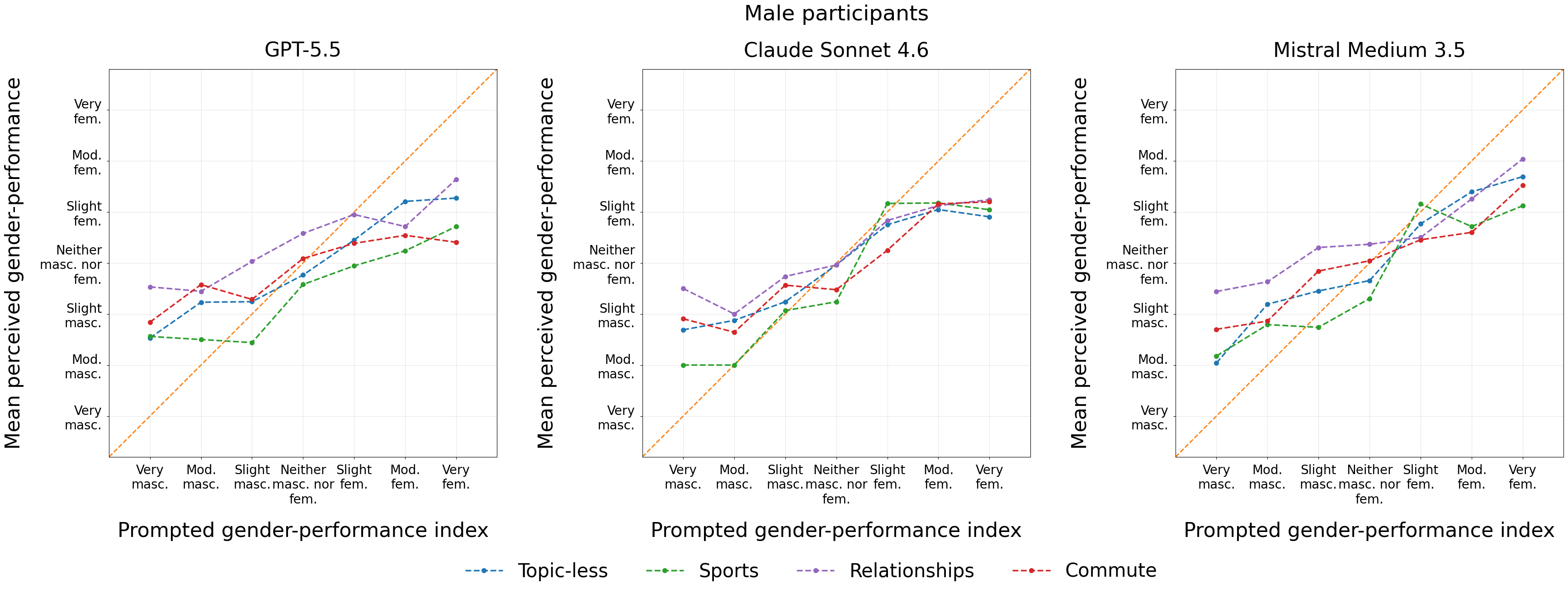}
        
    \end{minipage}

\caption{Model-specific comparisons of mean perceived gender across prompted gender-performance indices and conversation topics, separately by participants' self-reported gender. Within each participant-gender group, the three panels correspond to GPT-5.5, Claude Sonnet 4.6, and Mistral Medium 3.5, and coloured lines distinguish Topic-less, Sports, Relationships, and Commute conditions.}

    \label{fig:model_topic_by_participant_gender}

\end{figure}

Figure~\ref{fig:model_topic_by_participant_gender}
presents the relationship between assigned and
perceived gender separately for the three LLM models
and four conversation topics within each
participant-gender group.

% =====================================================
% FIGURE S10: MODEL-BY-TOPIC COMPARISONS
% =====================================================

% Page 1: female and male participants

\clearpage

\subsubsection{Gender-performance perceptions among gender minority participants}

\begin{figure}[H]
    \centering
    \includegraphics[
        width=0.75\textwidth,
        keepaspectratio
    ]{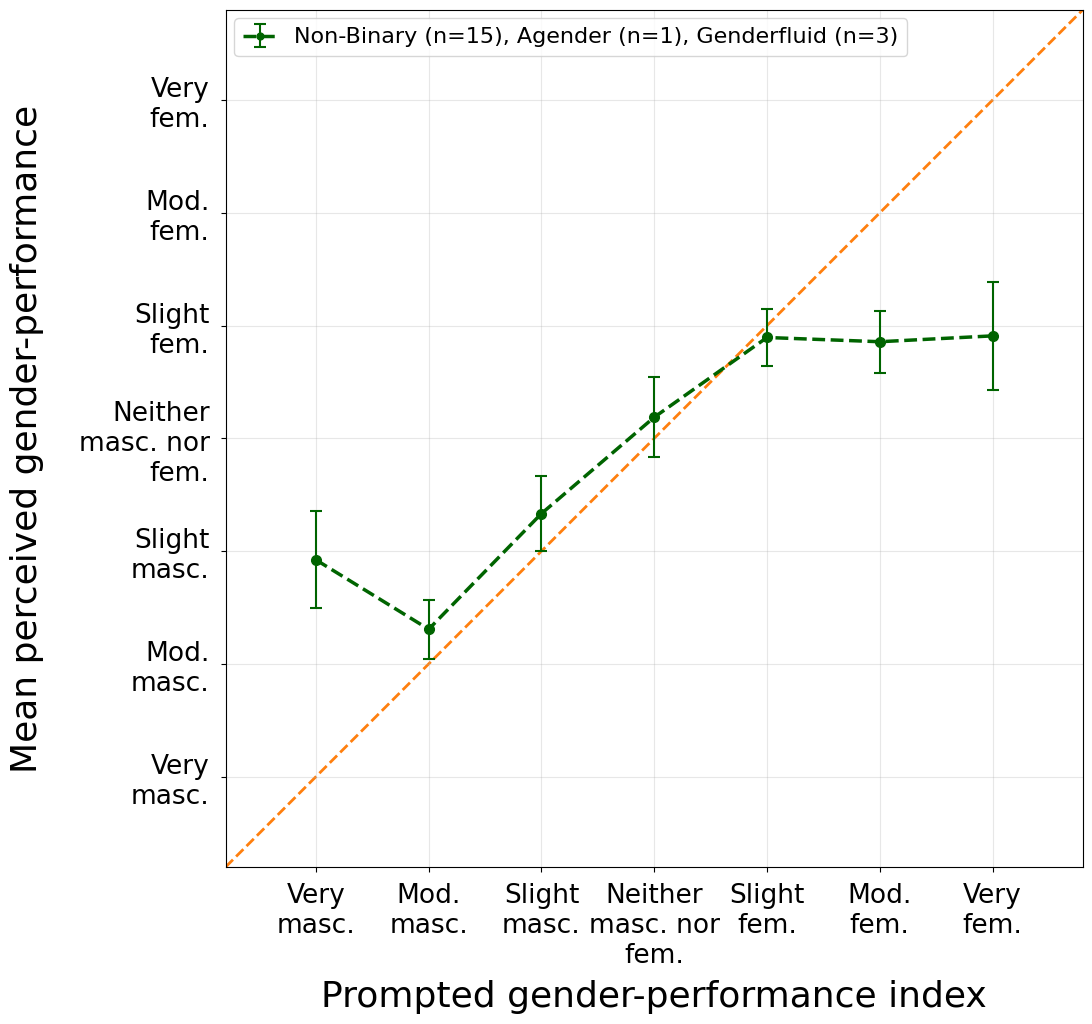}
    \caption{Mean perceived gender across assigned gender-performance indices among 19 participants identifying as non-binary, agender, or genderfluid. Points represent observed means, error bars indicate standard errors.}
    \label{fig:gender_minority_perception}
\end{figure}

Figure~\ref{fig:gender_minority_perception} presents mean perceived gender across assigned gender-performance indices for participants who identified as non-binary, agender, or genderfluid.

\clearpage
\printbibliography[title={References}]

\clearpage
\end{refsection}
\end{document}